\documentclass[a4paper]{article}

\usepackage{xparse,graphicx}
\usepackage[T1]{fontenc}
\usepackage[margin=1in]{geometry}
\usepackage{amsmath,amssymb}
\usepackage{amsthm}
\usepackage{mathtools}
\usepackage{caption,subcaption}
\usepackage{xurl}
\usepackage[hidelinks]{hyperref}
\usepackage{orcidlink}
\usepackage{thmtools}
\usepackage{thm-restate}
\usepackage[capitalize,noabbrev,nameinlink,sort&compress]{cleveref}

\newtheorem{theorem}{Theorem}[section]
\newtheorem{lemma}[theorem]{Lemma}
\newtheorem{definition}[theorem]{Definition}
\newtheorem{observation}[theorem]{Observation}
\newtheorem{remark}[theorem]{Remark}
\newtheorem{corollary}[theorem]{Corollary}
\newtheorem{hypothesis}[theorem]{Hypothesis}
\theoremstyle{definition}
\newtheorem{example}[theorem]{Example}
\AddToHook{env/example/begin}{\pushQED{\qed}}
\AddToHook{env/example/end}{\popQED}
\newtheorem{mixedintegerprogram}[theorem]{Mixed integer linear program}
\newtheorem{linearprogram}[theorem]{Linear program}
\theoremstyle{plain}

\crefname{constraint}{Constraint}{Constraints}
\creflabelformat{constraint}{(#2#1#3)}

\newcommand{\paperauthor}[4]{#1\if\relax\detokenize{#3}\relax\else\textsuperscript{\ensuremath{#3}}\fi \if\relax\detokenize{#4}\relax\else\,{\hypersetup{hidelinks}\orcidlink{#4}}\fi \if\relax\detokenize{#2}\relax\else\\{\small\hypersetup{hidelinks}\href{mailto:#2}{\nolinkurl{#2}}}\fi }
\makeatletter
\let\@author\@empty
\let\@funding\@empty
\renewcommand{\author}[5]{\ifx\@author\@empty\else\g@addto@macro\@author{\and}\fi \g@addto@macro\@author{\paperauthor{#1}{#2}{#3}{#4}}\if\relax\detokenize{#5}\relax\else
        \g@addto@macro\@funding{\noindent #1: #5\par}\fi
}
\let\@affiliations\@empty
\newcommand{\affiliation}[2]{\g@addto@macro\@affiliations{\textsuperscript{\ensuremath{#1}}#2\par}}
\renewcommand{\@maketitle}{\newpage
    \null
    \vskip 2em\begin{center}\let\footnote\thanks
        {\LARGE\@title\par}\vskip 1.5em{\large\lineskip .5em\begin{tabular}[t]{c}\@author\end{tabular}\par}\ifx\@affiliations\@empty\else
            \vskip 1em{\normalfont\normalsize\@affiliations}\fi
    \end{center}\ifx\@funding\@empty\else
        \begingroup
        \renewcommand{\thefootnote}{}\long\def\@makefntext##1{\noindent##1}\footnotetext[0]{\textbf{Funding:}\par\@funding}\endgroup
    \fi
    \par\vskip 1.5em}
\makeatother

\newenvironment{bracketenumerate}{\begin{enumerate}}{\end{enumerate}}

\makeatletter
\let\origsubsubsection\subsubsection
\RenewDocumentCommand{\subsubsection}{s o m}{\IfBooleanTF{#1}{\origsubsubsection*{#3\@addpunct{.}}}{\IfNoValueTF{#2}{\origsubsubsection[#3]{#3\@addpunct{.}}}{\origsubsubsection[#2]{#3\@addpunct{.}}}}}
\let\origparagraph\paragraph
\RenewDocumentCommand{\paragraph}{s o m}{\IfBooleanTF{#1}{\origparagraph*{#3\@addpunct{.}}}{\IfNoValueTF{#2}{\origparagraph[#3]{#3\@addpunct{.}}}{\origparagraph[#2]{#3\@addpunct{.}}}}}
\makeatother

\newcommand{\problemname}[1]{\textnormal{\textsc{#1}}}
\newcommand{\upsf}[1]{\textnormal{\textsf{#1}}}
\newcommand{\complclassfont}[1]{\upsf{#1}}
\newcommand{\np}{\complclassfont{NP}}
\newcommand{\pclass}{\complclassfont{P}}
\newcommand{\vcsp}[1][\Gamma]{\problemname{VCSP($#1$)}}
\newcommand{\csp}[1][\Gamma]{\problemname{CSP($#1$)}}
\newcommand{\ksat}[1][k]{\problemname{$#1$-Sat}}
\newcommand{\threeSat}{\ksat[3]{}}
\newcommand{\kcoloring}[1][k]{\problemname{$#1$-Coloring}}
\newcommand{\tw}{\upsf{tw}}
\newcommand{\pw}{\upsf{pw}}
\newcommand{\pwseth}{{pw-SETH}}
\newcommand{\twseth}{{tw-SETH}}
\newcommand{\arity}[1]{\upsf{arity}(#1)}
\newcommand{\costfun}[1]{\upsf{costf}(#1)}
\newcommand{\scope}[1]{\upsf{scope}(#1)}
\newcommand{\weight}[1]{\upsf{weight}(#1)}
\newcommand{\power}[1]{\upsf{pow}(#1)}
\newcommand{\Oh}{\mathcal{O}}
\newcommand{\Opt}{\mathsf{Opt}}
\newcommand{\cost}{\mathsf{cost}}
\newcommand{\additiveequiv}{\equiv_\Gamma}
\newcommand{\ccadditiveequiv}{q_\Gamma}
\newcommand{\moves}{\leadsto}
\newcommand{\ratN}{\mathbb{Q}}
\newcommand{\ratNnonneg}{\ratN_{\geq 0}}
\newcommand{\ratNpositive}{\ratN_{>0}}
\newcommand{\CC}{\mathcal{C}}
\newcommand{\eps}{\varepsilon} 

\title{Tight Algorithms for Conservative Finite-Valued CSPs Parameterized by Treewidth}

\author
    {Jakob Greilhuber}
    {jakob.greilhuber@cispa.de}{\star,\ddagger}{0009-0001-8796-6400}{}

\author
    {D\'aniel Marx}
    {marx@cispa.de}{\star}{0000-0002-5686-8314}{}

\author
  {George Osipov}
  {george.osipov@pm.me}
  {\dagger}
  {0000-0002-2884-9837}
  {Supported by the Swedish Research Council (VR) under grant 2024-00274.}

\author{Marta Piecyk}{marta.piecyk@cispa.de}{\star}{0000-0001-9162-8300}{}

\author
    {Paweł Rzążewski}
    {pawel.rzazewski@pw.edu.pl}{\S}{0000-0001-7696-3848}{Supported by the National Science Centre grant 2024/54/E/ST6/00094.}

\author{Stanislav \v{Z}ivn\'y}{standa.zivny@cs.ox.ac.uk}
{\|}
{0000-0002-0263-159X}
{Supported by UKRI EP/X024431/1.}

\affiliation{\star}{CISPA Helmholtz Center for Information Security, Saarbrücken, Germany}
\affiliation{\ddagger}{Saarland University, Saarbrücken, Germany}
\affiliation{\dagger}{Royal Holloway, University of London, UK}
\affiliation{\S}{Warsaw University of Technology, Warsaw, Poland}
\affiliation{\|}{Department of Computer Science, University of Oxford, UK}

\begin{document}

\maketitle
\thispagestyle{empty}

\begin{abstract}
    We consider valued constraint satisfaction problems (VCSPs) with a fixed finite domain $D$ and a fixed finite constraint language $\Gamma$, in the setting where $\Gamma$ contains all crisp unary constraints (called hard lists) and the remaining cost functions are finite-valued.
    We call such a language a \emph{finite-valued hard-list language}. Let $c_\Gamma$ be the size of the largest set $S\subseteq D$ such that $\Gamma$ can express
    (using the usual VCSP closure operations) a binary disequality function on elements of $S$.
    Our main result tightly connects the complexity of the problem parameterized by treewidth to this value $c_\Gamma$.
    \begin{enumerate}
        \item[(1)] The problem can be solved in time $(c_\Gamma)^{\mathsf{tw}} \cdot (|I| + |\mathcal{T}|)^{\Oh(1)}$ when given an instance $I$ with a tree decomposition $\mathcal{T}$ of width $\mathsf{tw}$ of the primal graph of $I$.
        \item[(2)] If $c_\Gamma \geq 2$, there is no such algorithm solving the problem in time $(c_\Gamma - \varepsilon)^{\mathsf{tw}} \cdot (|I| + |\mathcal{T}|)^{\Oh(1)}$ for any $\varepsilon > 0$, unless the Primal Treewidth SETH (and hence CNF-SETH) fails.
    \end{enumerate}
    In particular, this gives a far-reaching generalization of a similar result of Esmer et al. [ESA 2024] on the optimization version of list homomorphism for graphs.

    We say that $\Gamma$ is a finite-valued soft-list language if every cost function in $\Gamma$ is finite-valued and $\Gamma$ contains every unary $\{0,1\}$-valued cost function that takes value $1$ on exactly one domain element. A simple reduction shows that our result for the hard-list languages can be used to determine the value $c_\Gamma$ for soft-list languages. This exact characterization for soft-list languages can be seen as a refinement of
    the \textsf{P} vs.~\textsf{NP}-hard dichotomy result of Kolmogorov and {\v{Z}}ivn{\'{y}} [J. ACM 2013] for conservative finite-valued VCSP($\Gamma$), by distinguishing different levels of hardness among the \textsf{NP}-hard cases.

    The lower bound in our main result is fairly straightforward, as lower bounds for disequality are well understood.
    The algorithmic part uses LP-based preprocessing to reduce the number of domain elements available at the variables and a subtle decomposition of the domain that allows us to split the problem into independent subinstances.
    These subinstances are either polynomial-time solvable (using submodular function minimization) or can be solved using standard dynamic-programming approaches based on the fact that only a bounded number of domain elements are available at each variable. The main combinatorial challenge of the proof is showing that when preprocessing cannot make further progress and no further decomposition is possible, then
    the disequality function on a certain subdomain can be expressed, proving that the algorithmic result and the lower bound \mbox{tightly match}.

\end{abstract}

\clearpage

\section{Introduction}
\label{sec:intro}
\problemname{Constraint Satisfaction Problems} (\problemname{CSP}s) and their extensions provide a broad framework encompassing many fundamental algorithmic problems.
The \problemname{Valued Constraint Satisfaction Problem} (\problemname{VCSP}) extends this framework to optimization problems.
Let $D$ be a finite nonempty domain.
A cost function of arity $a$ is a function $f:D^a\to\ratNnonneg\cup\{\infty\}$.
For a fixed finite domain $D$ and a finite set $\Gamma$ of cost functions, an instance of \vcsp{} consists of a finite set $V$ of variables and a finite family $\CC$ of constraints.
Each constraint $C\in\CC$ is a triple $(f,S,w)$, where $f\in\Gamma$ is the cost function, denoted by $\costfun{C}$, $S\in V^{\arity{f}}$ is a tuple of $\arity{f}$ variables, and $w\in\ratNpositive$ is a positive rational weight.
The goal is to find an assignment of values to variables that minimizes the total weighted cost of all constraints in $\CC$.

Over the past decades, considerable effort has been devoted to characterizing
the sets $\Gamma$ for which \vcsp{} is polynomial-time solvable.
If every
$f\in\Gamma$ takes values in $\{0,\infty\}$, that is, if every constraint is
crisp, then \vcsp{} reduces to a decision problem whose goal is to find a
feasible assignment.
The Feder--Vardi Dichotomy Conjecture~\cite{FederV98}, proved by Bulatov~\cite{Bulatov17}
and Zhuk~\cite{Zhuk20} independently, states that every such problem is either polynomial-time solvable or
\np{}-hard.
For the case where every $f\in\Gamma$ is finite-valued, a dichotomy
theorem was proved by Thapper and {\v{Z}}ivn{\'{y}}
\cite{thapperComplexityFiniteValued2016}.
Finally, the combination of the work by Kozik and Ochremiak~\cite{KozikOchremiak15Algebraic} and by Kolmogorov, Krokhin, and
Rol\'inek~\cite{KKR17:sicomp} unified these results by establishing a dichotomy for every fixed finite constraint language $\Gamma$: \vcsp{} is either polynomial-time solvable or \np{}-hard.
The search for such dichotomy theorems can be viewed as a systematic search for all relevant algorithmic ideas: a dichotomy theorem identifies a collection of algorithmic techniques and shows that, when none of them applies, no further technique can yield a polynomial-time algorithm, unless $\pclass = \np$.

For each fixed $\Gamma$, the dichotomy proofs mentioned above either establish tractability or show that the problem is at least as hard as a known \np{}-hard problem, such as \kcoloring[3]{} or \problemname{Max Cut}.
This suffices to distinguish polynomial-time solvable cases from \np{}-hard ones, but does not provide a finer understanding of the complexity of the hard cases.
Intuitively, a problem equivalent to \kcoloring[100]{} should be considered harder than one equivalent to \kcoloring[3]{}: on $k$ variables, the former has a search space of size $100^k$, compared with $3^k$ for the latter.
This distinction is invisible from the perspective of polynomial-time solvability versus \np{}-hardness, since \kcoloring[100]{} and \kcoloring[3]{} are polynomial-time equivalent.
However, in some settings, differences of this kind can be established rigorously.
One such setting is the parameterized complexity of problems on graphs of bounded treewidth.

\paragraph{Treewidth} Treewidth is a combinatorial graph parameter that plays a central role in graph structure theory, most prominently in the Graph Minors project of Robertson and Seymour (see, e.g.,~\cite{RobertsonS86GraphMinorsII,RobertsonS86GraphMinorsV,RobertsonS95bGraphMinorsXIII}).
From an algorithmic perspective, treewidth measures how close a graph is to being a tree, and hence how effectively the bottom-up dynamic programming techniques used for trees can be extended to it.
Many applications in parameterized complexity, approximation-scheme design, and other areas rely crucially on the efficient solvability of problems on graphs of bounded treewidth~\cite{Courcelle90,BodlaenderK08,DemaineH08,CyganFKLMPPS15,Marx20FourShortStories}.
In particular, given a tree decomposition of width $\tw$, several fundamental problems can be solved in time $c^\tw\cdot n^{\Oh(1)}$ for some constant $c>1$ using fairly standard dynamic programming.
A natural goal is to determine the smallest possible value of $c$ for each problem.

A line of research initiated by Lokshtanov, Marx, and Saurabh~\cite{LokshtanovMS18} showed that many known algorithms are optimal in the following sense: a problem can be solved in time $c^\tw\cdot n^{\Oh(1)}$, but, assuming the Strong Exponential-Time Hypothesis (SETH)~\cite{ImpagliazzoP01,CalabroIP09}, no algorithm runs in time $(c-\eps)^\tw\cdot n^{\Oh(1)}$ for any $\eps>0$.
Recent preprints by OpenAI~\cite{OpenAI26ETH} and Claude (Anthropic)~\cite{ClaudeAnthropic26ETH} report refutations of the Exponential Time Hypothesis (ETH), and hence also refutations of SETH.
However, many of the lower bounds developed over the years remain valid under the weaker conjecture CNF-SETH --- that is, $n$-variable $m$-clause \textsc{Sat} with unbounded clause size cannot be solved in time $(2-\eps)^n\cdot (n+m)^{\Oh(1)}$ for any $\eps>0$ --- which has not been refuted as of this writing.
Prior work~\cite{IwataY15TWSETH,lampisPrimalPathwidthSETH,lampisTreewidthSETH}
introduced the even weaker Primal Pathwidth Strong Exponential Time Hypothesis (\pwseth{})
and Primal Treewidth Strong Exponential Time Hypothesis (\twseth{}), and showed that they also form a source of hardness for these lower bounds and are in some cases actually equivalent to them.
Both of these hypotheses follow from CNF-SETH.
For some problems, these lower bounds establish the optimality of straightforward dynamic programming, whereas for others, achieving the optimal base $c$ requires nontrivial algorithmic ideas.
For a broad range of tight treewidth and pathwidth bounds, including counting variants, and results on the hypotheses underlying these bounds, see~\cite{LokshtanovMS18,vanRooijEtAl18FastDP,CyganNPPRW22,IwataY15TWSETH,BorradaileLe16Domination,KatsikarelisLampisPaschos22Scattered,EgriMarxRzazewski18,CyganKratschNederlof18,CurticapeanLindzeyNederlof18,OkrasaPiecykRzazewski20,DubloisLampisPaschos21Mixed,MarxSankarSchepper21,DubloisLampisPaschos22Upper,FockeMarxRzazewski24,FockeEtAl25DominationAlgorithms,FockeEtAl25DominationHardness,LampisVasilakis24BoundedDegree,esmerListHomomorphismsDeleting2024ESA,lampisPrimalPathwidthSETH,Lampis26FiveShades,lampisTreewidthSETH}.
The primal graph (or Gaifman graph) of a \vcsp{} instance has the variables as its vertices, with two vertices adjacent if the corresponding variables occur together in some constraint.
Given a \vcsp{} instance and a tree decomposition of width $\tw$ of its primal graph, standard dynamic programming solves the instance in time $|D|^\tw\cdot n^{\Oh(1)}$.
Is this running time optimal for every $\Gamma$ for which \vcsp{} is \np{}-hard?
For some languages $\Gamma$, simple improvements are possible, for example when two values $a,b\in D$ play exactly the same role in every cost function of $\Gamma$.
This raises the possibility that less obvious algorithmic ideas could lead to further improvements.
Ideally, for every finite $\Gamma$ for which \vcsp{} is \np-hard, we would like to identify a constant $c_\Gamma>1$, give an algorithm running in time $c_\Gamma^\tw\cdot n^{\Oh(1)}$, and prove a conditional lower bound ruling out running time $(c_\Gamma-\eps)^\tw\cdot n^{\Oh(1)}$ for every $\eps>0$.

An important step toward the \problemname{CSP} and \problemname{VCSP} dichotomies was the resolution of their conservative cases~\cite{Bulatov11,KZ13}.
The term \emph{conservative} has been used with somewhat different meanings in the \problemname{CSP} and \problemname{VCSP} literature.
For the decision problem \csp{}, conservativity means that, for every subset $S\subseteq D$, a constraint restricting a variable to $S$ is available.
Equivalently, one may consider the list version of the problem, in which every variable $v$ is equipped with a ``hard list'' $L_v\subseteq D$, and the goal is to find a solution respecting all these lists.
In our setting, this means that $\Gamma$ contains every unary $\{0,\infty\}$-valued cost function.
In the \problemname{VCSP} literature, on the other hand, conservativity means that every finite-valued unary cost function is considered to be available.
Equivalently, for each $a\in D$, the unary cost function that takes value $1$ on $a$ and $0$ on $D\setminus\{a\}$ is available.
This can be interpreted as allowing a ``soft list'' for each variable, specifying a cost $c_{v,d}$ for assigning value $d$ to variable $v$.

\paragraph{Main result} Our main result determines $c_\Gamma$ when $\Gamma$ allows hard lists (that is, contains every unary $\{0,\infty\}$-valued cost function) and all its other cost functions are finite-valued.
We call such a language a \emph{finite-valued hard-list language}.
Let $\power{\Gamma}$ be the set of all cost functions that can be obtained from finite weighted sums of constraints from $\Gamma$ by minimizing over auxiliary variables and allowing an additive rational shift. Let $c_\Gamma$ be the size of the largest set $S\subseteq D$ such that $\power{\Gamma}$ contains the binary disequality function on $S$: that is, there is a function $g\in \power{\Gamma}$ such that $g(x,x)=1$ for every $x\in S$ and $g(x,y)=0$ for every $x,y\in S$, $x\neq y$. Note that in our context, we can always make $g$ have infinite or very large value outside $S$.

\begin{restatable}{theorem}{mainTheorem}\label{thm:main}
    Let $D$ be a fixed finite domain and $\Gamma$ a fixed finite-valued hard-list language over $D$.
    \begin{enumerate}
        \item[(1)] There is an algorithm that, given a \vcsp{} instance $I$ together with a tree decomposition $\mathcal T$ of its primal graph of width $\tw$, solves $I$ in time
              \[
                  (c_{\Gamma})^\tw\cdot (|I|+|\mathcal T|)^{\Oh(1)}.
              \]
        \item[(2)] If $c_\Gamma \geq 2$, for every $\eps>0$, there is no algorithm
              that solves an instance $I$ of \vcsp{}, given together with a tree decomposition $\mathcal T$ of its primal graph of width $\tw$, in time
              \[
                  (c_{\Gamma}-\eps)^\tw\cdot (|I|+|\mathcal T|)^{\Oh(1)},
              \]
              unless \twseth{} is false. The same lower bound holds under \pwseth{} with treewidth replaced by pathwidth.
    \end{enumerate}
\end{restatable}

As CNF-SETH implies both \twseth{} and \pwseth{}, these results are also true under CNF-SETH.

An example deferred to \cref{sec:gamma_pow_gamma_example} clarifies a delicate detail in the definition of $c_\Gamma$ and the statement of \cref{thm:main}.
It is not necessarily true that restricting every relation in $\Gamma$ to the set $S\subseteq D$ yields a language $\Gamma_S$ with the same complexity: in fact, there is a language $\Gamma$ such that even though $\vcsp{}$ requires $|S|^{\tw} \cdot n^{\Oh(1)}$ time for some $S \subseteq D$, for each $S' \subsetneq D$ the problem obtained by restricting $\Gamma$ to $S'$ is even polynomial-time solvable.
It is possible that in order to express the disequality function on $S$ appearing in $\power{\Gamma}$, we need to rely on values outside $S$. For example, it is possible that the optimum running time is $2^\tw\cdot n^{O(1)}$, but restricting the domain to every subset of size two gives a polynomial-time solvable problem.

The lower bound of \cref{thm:main}(2) follows fairly easily from the way $c_\Gamma$ is formulated. A binary disequality function over a domain $S\subseteq D$ of size $|S|\ge 3$ can be used to express \kcoloring[|S|]{}: replacing each edge with such a constraint, the optimum value is 0 if and only if the graph has a $|S|$-coloring.
It has been established in earlier work~\cite{lampisPrimalPathwidthSETH,lampisTreewidthSETH} that \kcoloring[q]{} for $q\ge 3$ has no $(q-\eps)^\tw\cdot n^{\Oh(1)}$ time algorithm for any $\eps>0$ (assuming \twseth{}, similarly for pathwidth under \pwseth{}).
For $|S|=2$, we can use a similar reduction from \textsc{Max Cut}, for which lower bounds are also known~\cite{lampisTreewidthSETH,lampisPrimalPathwidthSETH,IwataY15TWSETH,esmerListHomomorphismsDeleting2024ESA}.
Therefore, the majority of the proof is devoted to the upper bound of \cref{thm:main}(1): showing that there are algorithmic ideas that allow us to reduce the running time precisely to $(c_\Gamma)^\tw\cdot n^{\Oh(1)}$.
That is, we need to make a connection between $c_\Gamma$, which was defined by existence of cost functions in $\power{\Gamma}$, and the existence of algorithmically exploitable structural properties of \vcsp{}.

\paragraph{Soft lists} Let $\Gamma$ be a finite-valued language that allows all soft lists (that is, contains every unary $\{0,1\}$-valued cost function that takes value $1$ on exactly one element of $D$).
We call such a language a \emph{finite-valued soft-list language}.
Let $\Gamma'$ be the extension of $\Gamma$ to a finite-valued hard-list language by introducing every $\{0,\infty\}$-valued unary cost function.
It is easy to see that $\vcsp[\Gamma]$ and $\vcsp[\Gamma']$ have essentially the same complexity: hard lists can be simulated by imposing sufficiently large finite penalties on excluded values.
Thus, applying \cref{thm:main} to $\Gamma'$ yields a precise complexity characterization for every finite-valued soft-list language $\Gamma$.
In particular, our result refines the polynomial-time versus \np{}-hard dichotomy of Kolmogorov and {\v{Z}}ivn{\'{y}} \cite{KZ13} for finite-valued conservative \problemname{VCSP} by determining the optimal base of the exponent in the \np{}-hard cases.
We can therefore distinguish between the complexities of different \np{}-hard cases of \vcsp{}.

As we shall see, both the optimal algorithm and the value of $c_\Gamma$ are significantly easier to describe for finite-valued soft-list languages: in the NP-hard cases, $c_\Gamma$ is exactly the number of equivalence classes under a relation called additive equivalence.
Accordingly, we organize the proof by first establishing the result for finite-valued soft-list languages and then building on it to prove \cref{thm:main} for finite-valued hard-list languages.
Much of the technical difficulty in the hard-list setting lies in expressing all soft lists on a certain subset of the domain $D$.

\paragraph{Comparison with graph homomorphisms} A special case of \cref{thm:main} was established by Esmer et al.~\cite{esmerListHomomorphismsDeleting2024ESA} in their study of the \problemname{Edge-Deletion List Homomorphism} problem.\footnote{Their lower bound is stated under SETH; the corresponding bound under \twseth{} follows easily using their techniques.}
In our terminology, this corresponds to the case where, apart from the hard-list functions, $\Gamma$ contains only a single finite-valued cost function, which is binary, symmetric, and $\{0,1\}$-valued.
Even in this restricted setting, obtaining the optimal base of the exponent requires several nontrivial algorithmic ideas:

\begin{itemize}
    \item \textbf{Polynomial-time solvability via maximum flows.}
          A complete characterization must account for the polynomial-time solvable cases.
          These cases turn out to be reducible to maximum flow computations.
    \item \textbf{Incomparability.}
          If distinct values $a$ and $b$ both belong to the list of a variable and $a$ is always at least as good as $b$ (that is,  $a$ \emph{dominates} $b$), then $b$ can be removed from that list.
          Therefore, we may assume that the values allowed by the hard list of each variable are pairwise \emph{incomparable} (that is, no value dominates another).
          Consequently, the size of each list is bounded by the maximum number of pairwise incomparable values. This allows us to improve the running time of the standard dynamic programming algorithm by replacing the size of the domain with this potentially smaller bound in the base of the exponent.
    \item \textbf{Domain decomposition.}
          For certain cost functions, the domain $D$ can be partitioned into $D_1$ and $D_2$, allowing the problem to be reduced to separate instances over these two domains.
          To make this work, one must ensure that every list is contained entirely in either $D_1$ or $D_2$ and that variables with lists contained in different parts do not interact.
\end{itemize}

These algorithmic ideas must naturally be reflected in the proof of \cref{thm:main}.
Although they provide useful clues, their counterparts in the general setting take substantially different forms:

\begin{itemize}
    \item \textbf{Polynomial-time solvability.}
          The dichotomy theorem of Thapper and {\v{Z}}ivn{\'{y}} \cite{thapperComplexityFiniteValued2016} shows that the polynomial-time solvable cases of finite-valued \vcsp{} can be solved using a linear programming relaxation.
          For our result with hard lists, we use a reduction to submodular function minimization from the work of Cohen, Cooper, and Jeavons~\cite{CCJ08}, as was done in~\cite{KZ13}.

    \item \textbf{Noninteraction instead of incomparability.}
          Although incomparability (that is, no value dominates another) is a natural notion, it is not the right one in the general setting.
          For example, when all soft lists are available, any two distinct values are incomparable by definition.
          However, if two values play identical roles in every constraint of arity at least two and differ only in unary constraints, there is no need to retain both in the list of a variable.
          We therefore define the more general notion of \emph{noninteraction}, which turns out to be useful both for algorithmic preprocessing and for lower bounds.
    \item \textbf{Overlapping decompositions.}
          Rather than partitioning the domain into disjoint parts, we introduce decompositions of a different type.
          Roughly speaking, depending on $\Gamma$ there is a finite number of \emph{blocks}, which guide the decomposition process.
          The variables of the instance can then be assigned to these blocks, and our preprocessing ensures that only variables within the same block can be neighbors in the primal graph.
          This way, we decompose the input instance into a number of output instances that can be optimized separately.
          Interestingly, this is not a domain decomposition in the sense that variables assigned to different blocks can have the same domain value in their available sets (that is, not forbidden by hard-list constraints).

    \item \textbf{Local instance-specific preprocessing.}
          As noted above, if $a$ is at least as good as $b$ in every finite-valued constraint of $\Gamma$, then $b$ can be discarded whenever both values occur in the list of a variable.
          However, this dominance need not hold globally across all constraints in $\Gamma$.
          It may instead hold only for a particular variable $v$ in a particular instance: regardless of the values assigned to other variables, assigning $a$ to $v$ is always at least as good as assigning $b$.
          In this case, we can remove $b$ from the list of $v$.
          At the same time, the reverse may hold for another variable $v'$, allowing us to remove $a$ from its list.
          More generally, the alternative to $b$ may depend on the assignment to the neighbors of $v$.
          We use an LP-based technique to certify that values can be eliminated locally in this way.
          Related local preprocessing techniques restricting domain elements using LP have been studied in dead-end elimination (introduced in~\cite{DesmetEtAl92DEE})~\cite{goldsteinEfficientRotamer1994,lastersEnhancedDeadEnd1995}, for VCSPs with unary and binary cost functions~\cite{KosterVanHoeselKolen02}, and in the framework of relaxed improving mappings~\cite{shekhovtsovHigherOrderPersistency2016}.
          While earlier work used this type of preprocessing mainly as a heuristic way to improve running time, in our result this preprocessing provably improves the worst-case running time for certain constraint languages.
\end{itemize}

An interesting outcome of our study is that in the case of finite-valued soft-list languages, many of these ideas are unnecessary.
It is sufficient to consider a pairwise interacting set of values at each variable, and the maximum size of such a set is the optimal base for the \np-hard cases.
Therefore, no decomposition idea is relevant for the optimal algorithm and the local preprocessing is not needed.

\paragraph*{Article structure.}

We give a technical overview of the work in \cref{sec:tech_overview}.
\cref{sec:preliminaries} contains preliminaries, and \cref{sec:interaction_moves} contains central definitions and proofs necessary for both soft-list and hard-list languages.
The results for soft-list and hard-list languages are proven in \cref{sec:vcsp_soft_lists,sec:vcsp_hard_lists}, respectively.

\paragraph*{AI usage}
Our main result has been obtained from several conversations with AI tools, and we also used such tools for the purpose of writing and literature search, see \cref{sec:ai_usage} for more details.

\section{Technical Overview}
\label{sec:tech_overview}

In this section, we introduce the definitions necessary to discuss the technical results and sketch the main proofs.

A common theme in CSP dichotomy results is that, instead of considering $\Gamma$, it is more important to consider the set of all constraints that can be simulated using $\Gamma$: this closure of $\Gamma$ is a mathematically important object, which can be studied, for example, from the perspective of universal algebra. While our proof does not use universal algebra explicitly, we often need to consider cost functions that are not in $\Gamma$, but can be expressed as a gadget using constraints from $\Gamma$. We write $\power{\Gamma}$ for the expressive power~\cite{CCCJZ13,thapperComplexityFiniteValued2016} of $\Gamma$: cost functions obtained from finite weighted sums of constraints from $\Gamma$ by minimizing over auxiliary variables, allowing an additive rational shift provided that the resulting cost function is nonnegative.
It is easy to see that for any finite $\Delta\subseteq \power{\Gamma}$, an instance of $\vcsp[\Gamma\cup \Delta]$ can be reduced to an instance of $\vcsp{}$ by replacing each constraint using $\Delta$ by a constant-sized gadget that uses only cost functions from $\Gamma$. Moreover, this reduction increases treewidth only by at most a constant depending on $\Delta$. This means that in the lower-bound proof, we are essentially free to use constraints also from $\power{\Gamma}$.

\subsection{Soft-List Languages}
For soft-list languages, the optimum running time can be stated in a relatively simple way: after retaining one representative of each additive-equivalence class (defined below), an \np-hard problem is ``as hard as possible'': standard dynamic programming on the reduced domain has the optimal exponential base.
In this subsection, we formally state this result and sketch its proof, which introduces some important concepts in the study of \vcsp{}. In \cref{sec:hard_list_overview}, we extend the discussion to the hard-list case, where the lack of all unary cost functions makes further nontrivial preprocessing and decomposition techniques relevant.

\paragraph{Additive equivalence}
Let us focus first on the case where $\Gamma$ is a finite-valued soft-list language. The first observation is that in this case any two distinct elements are incomparable: for every pair of distinct elements $a,b\in D$, the language $\Gamma$ contains a unary constraint where $a$ is strictly better than $b$, as well as a unary constraint where $b$ is strictly better than $a$. At first, this suggests that we cannot obtain any significant improvement by restricting to an incomparable set of values.
However, we can still get some improvement if we look at the unary constraints and the constraints of higher arity separately.  Suppose that there are two values $a, b \in D$ that play exactly the same role in every cost function $f\in \Gamma$ of arity at least 2.
Then, at every variable $v$, it is sufficient to consider as a candidate value only one of $a$ and $b$: the one that has lower cost in the unary constraints on $v$.

Playing exactly the same role in every cost function $f\in \Gamma$ of arity at least 2 is an equivalence relation, and this observation implies that the maximum number of values we need to consider at a variable is the number of these equivalence classes. However, the situation is more complicated than that. Consider the following example,  where the soft-list family $\Gamma$ over $D=\{1,2,3\}$
consists of the following two functions and the soft lists:
\[
    \begin{aligned}
        f_1(x,y) & =
        \begin{cases}
            1 & \text{if } x=1 \text{ or } x=y=3, \\
            0 & \text{otherwise},
        \end{cases}
                 &
        f_2(x,y) & =
        \begin{cases}
            1 & \text{if } x=2 \text{ or } x=y=3, \\
            0 & \text{otherwise}.
        \end{cases}
    \end{aligned}
\]
Clearly, no two values of $D$ play the same role in these binary functions, partitioning $D$ into three singleton equivalence classes. However, if we define
$g(x,y)$ to be 1 if $x=y=3$ and 0 otherwise,
then  $f_i(x,y)$ can be expressed as $g(x,y)$ plus a unary function that is 1 if $x=i$ and 0 otherwise. Thus every instance of $\vcsp{}$ can be rewritten with constraints involving $g$ and unary constraints. As $1$ and $2$ play the same role in the function $g$, this means that we have only two equivalence classes: $\{1,2\}$ and $\{3\}$.

This example suggests that we should consider two values equivalent if they can be made equivalent in every cost function by adding unary constraints. We formalize this in the following definition: we say that $a,b\in D$ are \emph{additively equivalent} in $\Gamma$ if for every $f\in \Gamma$ of arity $r$ and every $i\in [r]$, there is a constant $c^{a,b}_{f,i}$ such that if an $r$-tuple has value $b$ at position $i$, changing that value to $a$ adds $c^{a,b}_{f,i}$ to its $f$-value, independently of the other entries. In the example above, we have $c^{1,2}_{f_1,1}=1$ and $c^{1,2}_{f_2,1}=-1$. It
is easy to show that additive equivalence is transitive and forms an equivalence relation $\additiveequiv$. It follows that essentially we have to consider only one representative from each class, or, in other words, we can reduce the problem to an instance of \vcsp[\Gamma']{} where the domain size is the number of $\additiveequiv$ classes in $\Gamma$.
This results in an
algorithm where the base of the exponent is the number of such equivalence classes.
A similar use of equivalence classes underlies the backtracking algorithm of~\cite{HelaouiNaanaa13} for \vcsp{} with unary and binary constraints.
It turns out that in the case of soft-list languages, this algorithmic idea already gives optimal running time.

\begin{theorem}[Follows from~\cref{thm:soft_lists_alg,thm:soft_lists_reversing_move,thm:soft_lists_base_equality,thm:lb_from_c_gamma}]
    \label{thm:soft_lists_intro}
    Let $D$ be a fixed finite domain and $\Gamma$ a finite-valued soft-list language over $D$.
    Let $\ccadditiveequiv$ be the number of equivalence classes of $\additiveequiv$.
    Then, the following hold:
    \begin{enumerate}
        \item[(1)]
              There is an algorithm that, given a \vcsp{} instance $I$ together with a tree decomposition $\mathcal T$ of its primal graph of width $\tw$, solves $I$ in time
              \[(\ccadditiveequiv)^{\tw}\cdot (|I|+|\mathcal T|)^{\Oh(1)}.\]

        \item[(2)] If \vcsp{} is \np-hard, then for every $\eps>0$,
              no algorithm can solve the problem in time
              \[(\ccadditiveequiv-\eps)^{\tw}\cdot (|I|+|\mathcal T|)^{\Oh(1)},\] when the input instance $I$ is given together with a tree decomposition $\mathcal{T}$ of width $\tw$ of the primal graph of $I$,
              unless \twseth{} is false.
    \end{enumerate}
\end{theorem}
The algorithmic result (1) follows essentially from the discussion above.
For the lower bound (2), we show that the number $\ccadditiveequiv$ of equivalence classes is the same as $c_\Gamma$. Therefore, the same straightforward lower bound arguments as in the lower bound of \cref{thm:main} (expressing \kcoloring[c_\Gamma]{}  or \textsc{Max Cut})  can be used.

\begin{restatable}{lemma}{thmSoftListBaseEquality}
    \label{thm:soft_lists_base_equality}
    Let $D$ be a fixed finite domain, and $\Gamma$ be a finite-valued soft-list language over $D$.
    If $c_{\Gamma} \geq 2$, then $c_{\Gamma} = \ccadditiveequiv$.
\end{restatable}

It is easy to see that $c_\Gamma \leq \ccadditiveequiv$: if $a$ and $b$ are additively equivalent, then we cannot express a disequality on $a$ and $b$. The difficult part of the proof is showing $c_\Gamma \ge \ccadditiveequiv$: if we take one value from each equivalence class, then we can express a disequality on the set $S$ of these values. In fact, we show a stronger statement: any finite cost function on $S$ can be expressed. While this might seem like overkill for the lower bound (as discussed above, it is sufficient to express a disequality), a general  proof for all functions is not significantly more complicated than a proof for a fixed function such as disequality.
Moreover, in the hard-list case, we indeed need to express more complicated cost functions in an inductive step, and hence later in the overview we will refer back to this construction. For simplicity, we state the result here for the case when the elements of $D$ are pairwise additively nonequivalent.

\begin{lemma}[Follows from \cref{thm:soft_lists_reversing_move,thm:soft_lists_express_costs}]
    \label{thm:soft-all-cost-intro}
    Let $D$ be a finite domain and $\Gamma$ a finite, finite-valued soft-list language over $D$ whose domain elements are pairwise additively nonequivalent.
    If \vcsp{} is not polynomial-time solvable, then $\power{\Gamma}$ contains every finite-valued cost function.
\end{lemma}

\paragraph{Interaction}
To prove \cref{thm:soft-all-cost-intro}, we need to understand what it means for the values in a set to be pairwise additively nonequivalent.
The lower-bound constructions crucially depend on the following definition, both for soft-list and hard-list languages. Because we are using this definition in the case of hard-list languages, we require that neither value in each of the two pairs dominates the other. For soft-list languages, this condition holds automatically, as the unary cost functions make the values pairwise incomparable. However, this extra condition will play a role for hard-list languages.

\begin{restatable}[Interacting pairs]{definition}{interactingPairs}
    \label{def:interaction}
    Let $D$ be a finite domain and $\Gamma$ a constraint language over $D$.
    Distinct elements $a,b\in D$ form an \emph{interacting pair} if neither dominates the other and there are distinct elements $c,d\in D$, neither of which dominates the other, and a finite-valued binary cost function $f\in\power{\Gamma}$ such that
    \[
        f(a,c)-f(b,c)\neq f(a,d)-f(b,d).
    \]
    We say that $\{c,d\}$ and $f$ \emph{witness} the interaction of $a,b$.
    A set is \emph{pairwise interacting} if every pair of its distinct elements interacts.
    If two distinct elements do not interact, they form a \emph{noninteracting pair}.
\end{restatable}
Intuitively, the inequality means that if we consider the function $f(x,y)$ and look at the unary functions  $f_c(x)$ and $f_d(x)$ obtained by substituting $y=c$ and $y=d$, respectively, then these two unary functions restricted to $\{a,b\}$ are really different and not just by a constant shift: replacing $a$ with $b$ changes the cost in different ways in the two unary functions. Note that the inequality is equivalent to $f(a,c)-f(a,d)\neq f(b,c)-f(b,d)$, giving also the interpretation that substituting $x=a$ and $x=b$ leads to different unary functions. Therefore, in a sense, the function $f$ can distinguish between $a$ and $b$ and can propagate the choice between them: depending on this choice at $x$, the resulting unary constraint on $y$ changes.
A very similar definition is used by Helaoui and Naanaa~\cite{HelaouiNaanaa13} in the context of modularity.

For soft-list languages, additive equivalence is exactly the same as not interacting. Assuming that the elements of $D$ are pairwise not additively equivalent, we may assume that the elements of $D$ are pairwise interacting.
\begin{lemma}[Follows from \cref{thm:additive_noninteraction}]
    \label{thm:additive_noninteraction_intro}
    Let $D$ be a finite domain and $\Gamma$ be a finite-valued soft-list language over $D$.
    Then, two distinct elements $d_1,d_2 \in D$ are additively equivalent if and only if they do not interact.
\end{lemma}

\paragraph{Moves}
The definition of interaction means that the choice between $a$ and $b$ at variable $x$ has some kind of effect on how $c$ and $d$ are valued at variable $y$. When transferring information in a lower-bound proof, we would like to make this effect more specific: ideally, we would like to say that $x=a$  forces $y=c$, while $x=b$ forces $y=d$. The following definition expresses the existence of such a binary function:

\begin{restatable}[Moves and reversing moves]{definition}{movesAndReversingMoves}
    \label{def:move}
    \label{def:reversing_move}
    Let $D$ be a domain and $\Gamma$ a constraint language over $D$.
    Let $\{a,b\}$ and $\{c,d\}$ be two subsets of $D$ of size $2$ each.
    A binary cost function $g$ is a \emph{move} from $(a,b)$ to $(c,d)$ if
    \[
        g(a,c)=g(b,d)<g(a,d)=g(b,c).
    \]
    We call the move \emph{finite-valued} if $g(x,y)<\infty$ for all $x,y\in D$.
    If there is a move from $(a,b)$ to $(c,d)$ in $\power{\Gamma}$ we write $(a,b) \moves (c,d)$.
    If there is a move from $(a,b)$ to $(c,d)$ or to $(d,c)$ we write $\{a,b\} \moves \{c,d\}$.

    A move from $(a,b)$ to $(b,a)$ is a \emph{reversing move on $\{a,b\}$}; equivalently, it satisfies
    \[
        g(a,b)=g(b,a)<g(a,a)=g(b,b).
    \]
    We say that $\Gamma$ expresses a reversing move on $\{a,b\}$ if such a function belongs to $\power{\Gamma}$.
\end{restatable}

Intuitively, an expressible reversing move on $\{a,b\}$ is a function $f\in\power{\Gamma}$ that expresses the binary disequality function: $f(a,b)=f(b,a)<f(a,a)=f(b,b)$. The dichotomy result of Thapper and {\v{Z}}ivn{\'{y}} \cite{thapperComplexityFiniteValued2016} for soft-list \vcsp{} implies that
if \vcsp{} is not polynomial-time solvable, then there are distinct $a,b\in D$ such that there is a reversing move on $\{a,b\}$. Moreover, if there is such a reversing move, then in a fairly standard way we can express every finite-valued cost function over $\{a,b\}$.

To prove \cref{thm:soft-all-cost-intro}, we need to express finite-valued cost functions not only on $\{a,b\}$, but on the full domain. To make this possible, we need to ``translate'' other elements of $D$ into a combination of values in $\{a,b\}$. The key combinatorial insight that makes this possible in the case of soft-list languages (but need not hold for hard-list languages) is that there is a move from any interacting pair to $\{a,b\}$.
\begin{restatable}{lemma}{interactingPairsMove}
    \label{thm:interacting_pairs_move}
    Let $D$ be a finite domain and $\Gamma$ a constraint language over $D$ that expresses all soft lists.
    For any interacting pairs $\{a,b\}$ and $\{c,d\}$, there is a finite-valued move $\{a,b\} \moves \{c,d\}$.
\end{restatable}
The proof of \cref{thm:interacting_pairs_move} turns out to be a relatively simple case analysis. The definition of interacting pairs means that there is a pair $\{a',b'\}$ for $\{a,b\}$ and a pair $\{c',d'\}$ for $\{c,d\}$ satisfying the conditions in \cref{def:interaction}. Then we can consider different cases depending on, for example, whether $\{a,b\}$ and $\{c',d'\}$ satisfy \cref{def:interaction} to obtain a sequence of moves from $\{a,b\}$ to $\{c,d\}$ and use the fact that moves are transitive in the case of soft-list or hard-list languages.

\paragraph{Translation to a 2-element set}
The proof of \cref{thm:soft-all-cost-intro} is based on a translation principle used already by Esmer et al.~\cite{esmerListHomomorphismsDeleting2024ESA}.
As discussed above, the assumption that \vcsp{} is not polynomial-time solvable implies that there is a reversing move on some $\{a,b\}$ and $\power{\Gamma}$ contains every finite-valued cost function on $\{a,b\}$. We construct a gadget as follows.
Consider an ``input'' variable $x$ and, for every $\{c,d\}\subseteq D$ of size two, an ``output'' variable $y_{\{c,d\}}$.
For every such $\{c,d\}$, let us introduce a constraint on $x$ and $y_{\{c,d\}}$ whose cost function is the move from $\{c,d\}$ to $\{a,b\}$. Let us add unary cost functions on $y_{\{c,d\}}$ such that any value other than $a$ or $b$ has prohibitively large cost. Consider now assigning a value $c$ to $x$. It has some number of minimum-cost ways of extending it to the $y$ variables; that is, for every $c$ there is some number of vectors in $\{a,b\}^{\binom{|D|}{2}}$ that result in minimum cost. The key observation is that for distinct $c,d\in D$, these vectors are distinct. Indeed, consider the variable $y_{\{c,d\}}$ and recall that a constraint with a move from $\{c,d\}$ to $\{a,b\}$ was introduced on $x$ and $y_{\{c,d\}}$. Therefore, the value of this variable in any vector corresponding to $c$ should be different from the value of this variable in any vector corresponding to $d$. By introducing additional unary cost constraints on variable $x$, we can ensure that for any $c\in D$ on $x$, the minimum-cost extension has the same cost. This means that we can express a cost function where in every minimum-cost assignment, the combination of $a$'s and $b$'s on the $\binom{|D|}{2}$ variables $y$ uniquely determines the value of variable $x$. This last step, introducing the normalizing unary cost constraints on $x$, may seem like a minor technical step. However, as we shall see, this step plays an important role in the difference between the soft-list case and the hard-list case, as in the latter case adding normalizing unary cost constraints may not always be possible.

To express a cost function $f$ of arity $r$ required by \cref{thm:soft-all-cost-intro}, we introduce $r$ copies of this construction. This results in a cost function where
in every minimum-cost assignment, the combination of $a$'s and $b$'s on the $r\binom{|D|}{2}$ variables $y$ uniquely determines the values of the $r$ variables $x$.
Now we can introduce a constraint of arity $r\binom{|D|}{2}$ on all the $y$ variables that gives a specific cost for each combination in $\{a,b\}^{r\binom{|D|}{2}}$. We set this cost depending on the assignment on the $x$ variables forced by the value of the $y$ variables, and on what cost $f$ assigns to this combination in $D^r$. As we assumed that every finite-valued cost function on $\{a,b\}$ is in $\power{\Gamma}$, this construction proves that $f$ is also in $\power{\Gamma}$.

\begin{figure}[!t]
    \centering
    \includegraphics{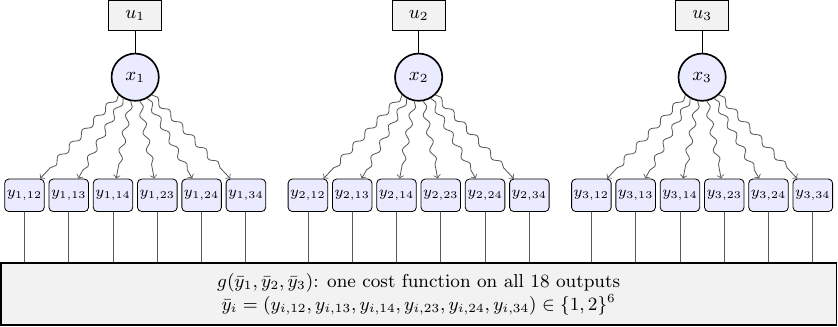}
    \caption{Translation construction for $d=|D|=4$ and $r=3$, with
        $D=\{1,2,3,4\}$ and a reversing pair $\{1,2\}\subseteq D$.
        Each input $x_i\in D$ is connected to six outputs:
        the edge to $y_{i,jk}$ represents a move from $\{j,k\}$ to $\{1,2\}$.
        Unary penalties restrict each output to $\{1,2\}$ (omitted from the drawing),
        and the unary cost $u_i$ normalizes the minimum extension cost independently of $x_i$.
        Each minimizing output tuple $\bar y_i$ determines $x_i$ uniquely.
        After sufficiently weighting the move gadgets, the cost function $g$ on the
        $3\binom{4}{2}=18$ outputs assigns the desired cost $f(x_1,x_2,x_3)$ to the corresponding
        encodings and a sufficiently large cost to all other tuples.
        Minimizing over the outputs realizes $f$ up to an additive constant.}
    \label{fig:translation-four-three}
\end{figure}

\cref{fig:translation-four-three} illustrates this construction for a domain of size four and a ternary cost function.

\subsection{Hard-List Languages}
\label{sec:hard_list_overview}

In the case of hard-list languages, each variable has a list $L_v$ of \emph{available elements,} which are the values not excluded by an infinite cost in a unary constraint.
An easy observation is that the base of the exponent in the running time can be improved from the domain size to the (potentially much smaller) maximum list size: standard dynamic programming on a tree decomposition needs to consider only assignments that use an available element at each variable in a bag \cite{freuderComplexityKTree1990,KosterVanHoeselKolen02,terriouxBoundedBacktracking2003}. Therefore, our main algorithmic goal in this section is to argue that the maximum list size can be provably decreased for some hard-list languages.

\begin{restatable}{lemma}{availableSetsDP}
    \label{thm:available_sets_dp}
    Let $D$ be a fixed finite domain and $\Gamma$ a fixed finite constraint language over $D$ whose cost functions are finite-valued or hard lists.
    Given an instance $I$ of \vcsp{} with variable set $V$ and a tree decomposition $\mathcal T$ of its primal graph of width $\tw$, set $c=\max\{|L_v|:v\in V\}$.
    One can compute $\Opt(I)$ and an optimal assignment in time
    $
        c^{\tw}\cdot (|I|+|\mathcal T|)^{\Oh(1)}.
    $
\end{restatable}

The fundamental difference between soft-list and hard-list languages is that in the latter case, \cref{thm:interacting_pairs_move} is no longer true: it may be the case that for some interacting pairs $\{a,b\}$ and $\{c,d\}$, there is no finite-valued move from $\{a,b\}$ to $\{c,d\}$.
This allows us to identify possibly overlapping blocks of the domain that do not interact with each other, and to reduce the problem to separate instances over these blocks.

\paragraph{The move graph}
As there may be no finite-valued move between certain interacting pairs, it is natural to define a graph that expresses which moves exist; this idea was also used in~\cite{esmerListHomomorphismsDeleting2024ESA}.
The \emph{move graph} $G_\Gamma$ (of $\Gamma$) has a vertex $(a,b)$ for each \emph{ordered} pair $a,b\in D$ such that $\{a,b\}$ is an interacting pair. There is an edge between $(a,b)$ and $(c,d)$ if there is a finite-valued move $(a,b) \moves (c,d)$.
A similar graph appeared in~\cite{Takhanov10,KZ13,DvorakKolmogorov25}.

In the case of soft-list languages, the move graph is complete (assuming the elements are pairwise additively nonequivalent and a reversing move exists). This is no longer true for hard-list languages, and in fact the move graph may have multiple connected components (see \cref{ex:move_graph}). A key insight in the hard-list case is that we can essentially treat each connected component of the move graph separately, and that the optimal base of the exponent is determined by the ``hardest'' component.

\clearpage
\begin{example}\label{ex:move_graph}
    Let $D=\{a,b,c,d\}$ and let $\Gamma$ consist of all hard lists and the symmetric binary cost function
    \[
        \begin{array}{c|rrrr}
            f & a & b & c & d \\\hline
            a & 2 & 1 & 1 & 2 \\
            b & 1 & 2 & 1 & 2 \\
            c & 1 & 1 & 2 & 3 \\
            d & 2 & 2 & 3 & 2
        \end{array}.
    \]
    Both $a$ and $b$ dominate $d$, while the four pairs
    $\{a,c\}$, $\{a,b\}$, $\{c,b\}$, and $\{c,d\}$ are interacting, each witnessed by itself and $f$.
    The move graph (\cref{def:move_graph}) has exactly three connected components:
    \[
        \begin{aligned}
            C_1 & =\{(a,c),(c,a),(a,b),(b,a),(c,b),(b,c)\}, \\
            C_2 & =\{(c,d)\},
            \qquad
            C_3=\{(d,c)\}.
        \end{aligned}
    \]
    Indeed, on $S=\{a,b,c\}$ the function $f$ has cost $2$ on equal arguments and $1$ on distinct arguments.
    It therefore gives reversing moves on every pair in $S$, and interaction between these pairs connects all six orderings in $C_1$.
    On $\{c,d\}$, $f$ gives a loop at each of $(c,d)$ and $(d,c)$, but no expressible finite-valued move connects these two vertices.
\end{example}

It is not difficult to observe that every component of the move graph is a clique: if $(a,b)$ and $(c,d)$ are in the same component, then there is a path of moves from $(a,b)$ to $(c,d)$, and by transitivity of moves, there is a move from $(a,b)$ to $(c,d)$. Therefore, if a component contains both $(a,b)$ and $(b,a)$ (as in the case of $C_1$), then there is a reversing move on $\{a,b\}$, which will be needed to prove hardness using the values appearing in the component.
We call such a component a \emph{reversing block,} and we  will focus on such blocks in the hardness proof.
It can be observed that if $(a,b)\moves (c,d)$, then $(b,a)\moves (d,c)$. Therefore, if a component $C$ does not contain a reversing move, then reversing every pair of $C$ gives another component $C'$ (see, e.g., the components $C_2$ and $C_3$ in the example above). In this case, we call the union of $C$ and $C'$ a \emph{nonreversing block.}

An interacting pair $\{a,b\}$ \emph{belongs} to a block $B$ if $(a,b)\in B$; since blocks are closed under reversal, this is equivalent to $(b,a)\in B$.
More generally, a pairwise interacting set $S\subseteq D$ with $|S|\geq2$ \emph{belongs} to $B$ if every pair of distinct elements of $S$ belongs to $B$.
The following important lemma shows that every pairwise interacting set of size at least two belongs to a unique block.

\begin{restatable}{lemma}{availableSetBlock}
    \label{thm:available_set_block}
    Let $D$ be a finite domain and $\Gamma$ a finite, finite-valued hard-list language over $D$.
    Let $L\subseteq D$ be pairwise interacting and of size at least $2$.
    All pairs of distinct elements of $L$ belong to one block of $G_\Gamma$.
\end{restatable}

\paragraph{Preprocessing}
The definitions of interaction and blocks of the move graph allow the following three preprocessing rules, which are used in the proof of \cref{thm:main} to reduce the instance to a simpler form.
\begin{itemize}
    \item \textbf{Dominated values.} If a value $a$ dominates another value $b$, then $b$ can be removed from the list of any variable where both $a$ and $b$ are available without changing the optimum value of the instance. Therefore, we can assume that no value is dominated by another value in the same list.
    \item \textbf{Noninteracting values.} After removing dominated values, if two values $a$ and $b$ available at $v$ do not interact, then it is sufficient to keep only one of them: the other one plays
          the same role in every constraint of arity at least two, up to an additive difference in the cost.
          Therefore, if we keep only the one with lower total cost after summing these additive differences over all incident constraints, we can ignore the other value without changing the optimum value of the instance. This means that we can assume that all values in the lists are pairwise interacting. In particular, it follows by \cref{thm:available_set_block} that, for any variable $v$ with at least two available values, there is a block $B_v$ of the move graph such that any pair of available values at $v$ is in $B_v$.
    \item \textbf{Decomposing the instance.} Let $V_B$ be the set of variables $v$ with $B_v=B$. We claim that we can separate the instance into subinstances on the variable sets $V_B$ for each block $B$ of the move graph. Suppose that there is a binary constraint $f(x,y)$ with $x\in V_B$ and $y\in V_{B'}$ for distinct blocks $B$ and $B'$. Suppose that $a,b$ are available at $x$ and $c,d$ are available at $y$. Then $\{a,b\}$ belongs to $B$ and $\{c,d\}$ belongs to $B'$, and by definition of the move graph, $\{a,b\}$ and $\{c,d\}$ do not interact. This means that the constraint $f(x,y)$ can be expressed as a sum of unary constraints on $x$ and $y$, and therefore can be replaced by these unary constraints without changing the optimum value. The argument can be extended to higher-arity constraints.
\end{itemize}

Suppose that these preprocessing steps create a subinstance where every nonsingleton available set belongs to the same block $B$; let $k_B$ be the maximum size of an available set in this subinstance. If this block $B$ is nonreversing, then earlier work implies that this subinstance can be solved in polynomial time~\cite{CCJ08,KZ13,thapperComplexityFiniteValued2016}. Otherwise, we can solve the subinstance using a standard dynamic programming approach in time $(k_B)^\tw\cdot n^{\Oh(1)}$: the maximum number of values that we need to consider at each variable is at most $k_B$.

By analogy to the soft-list case, it may be tempting to conjecture that this is the optimum approach: the best possible base of the exponent is the maximum size of a pairwise interacting set $D_B$ belonging to any reversing block $B$ of the move graph. However, if we try to follow the proof of the soft-list case but restricted to this block $B$, we run into a difficulty at the very end of the proof. It is true that there are $a,b\in D_B$ such that there is a reversing move on $\{a,b\}$, and therefore every finite-valued cost function on $\{a,b\}$ can be expressed. It is also true that for every two distinct elements $c,d\in D_B$ there is a finite-valued move from $\{a,b\}$ to $\{c,d\}$. The problem is that when we put these moves together to express an arbitrary cost function over $D_B$, we need to add unary cost functions on the $x$ variables to normalize their cost. In our case, it may not be true that every finite-valued unary cost function on $D_B$ is expressible.

\paragraph{Expressive sets.}
Hence, we consider subsets of values where all unary cost functions are available.
For $D' \subseteq D$ we say that $\Gamma$ expresses all soft lists on $D'$ if, for each $B \subseteq D'$ there is a finite-valued unary function $g \in \power{\Gamma}$ and a constant $c$ such that $g(d) = c$ for $d \in B$, and $g(d) = c + 1$ for $d \in D' \setminus B$.

\begin{restatable}[Expressive set]{definition}{expressiveSet}\label{def:expressive_set_intro}\label{def:expressive_set}
    Let $D$ be a finite domain and $\Gamma$ a finite, finite-valued hard-list language over $D$.
    A set $D' \subseteq D$ is \emph{expressive} if
    \begin{bracketenumerate}
        \item $D'$ is pairwise interacting, and
        \item $\Gamma$ expresses all soft lists on $D'$.
    \end{bracketenumerate}
\end{restatable}

With these definitions at hand, we can state our main results for hard lists in a different way.
Concretely, we show that for all finite-valued hard-list languages, the constant $c_\Gamma$ is equal to the maximum size of an expressive set $D_B\subseteq D$ that belongs to a reversing block $B$, and one if there is no such set $D_B$.

\begin{restatable}{lemma}{thmHardListBaseEquality}
    \label{thm:hard_lists_base_equality}
    Let $D$ be a finite domain, and $\Gamma$ a finite, finite-valued hard-list language over $D$.
    Then $c_\Gamma$ is equal to \[
        \max\bigl(\{1\}\cup
        \{|S| \mid S\subseteq D\text{ is expressive and belongs to a reversing block of }G_\Gamma\}\bigr).
    \]
\end{restatable}

In \cref{thm:hard_lists_base_equality}, the upper bound on $c_\Gamma$ follows easily:
if disequality is defined on a set $S\subseteq D$ with $|S| \geq 2$, then $S$ is expressive and belongs to a reversing block of the move graph.
For the lower bound, we can follow the same arguments as in \cref{thm:soft-all-cost-intro} for soft-list languages: if we assume that $S$ is an expressive set, then the availability of the finite-valued unary cost functions makes the proof go through.

We can prove the upper bound of \cref{thm:main}(1) by ensuring that every available set $L_v$ is expressive and belongs to a reversing block.
Then by \cref{thm:hard_lists_base_equality}, the size of the largest available set is at most $c_\Gamma$, and we can apply \cref{thm:available_sets_dp} to solve the instance in time $(c_\Gamma)^\tw\cdot n^{\Oh(1)}$. Thus the rest of the overview is devoted to presenting the preprocessing techniques that ensure that every available set is expressive and belongs to a reversing block.

\paragraph{Locally necessary values.}
Consider an instance $I$ of \vcsp{}. We say that a value $a\in L_v$ is \emph{locally necessary} at variable $v$ if there is a partial assignment of the variables $V\setminus \{v\}$ respecting their available sets such that $a$ is the unique best extension of this assignment to a full assignment of $I$.
If $a$ is \emph{not} locally necessary at $v$, then we can exclude it from the available set of $v$ without changing the optimum cost: by definition, no matter what the values of the other variables in an optimum assignment are, there is a value other than $a$ that can optimally extend this partial assignment to $v$.

\begin{example}\label{ex:local_necessity_single_block}
    Let $D=\{a,b,c\}$ and let $\Gamma$ consist of all hard lists and the symmetric binary cost function
    \[
        \begin{array}{c|rrr}
            f & a & b & c \\\hline
            a & 0 & 1 & 2 \\
            b & 1 & 4 & 1 \\
            c & 2 & 1 & 0
        \end{array}.
    \]
    No two domain elements dominate one another, and each pair $\{x,y\}$ witnesses its own interaction, since
    $f(x,x)+f(y,y)\neq 2f(x,y)$.
    Thus $D$ is pairwise interacting, and all pairs belong to one block by \cref{thm:available_set_block}.
    This block is reversing: pinning auxiliary variables to $c$ expresses
    \[
        g(x,y)=f(x,y)+2f(x,c)+2f(y,c),
    \]
    for which $g(a,b)=g(b,a)=7<8=g(a,a)=g(b,b)$.

    Nevertheless, in any instance over $\Gamma$, the value $b$ is not locally necessary at a variable whose available set is $D$.
    Indeed, for every $z\in D$,
    \[
        2f(b,z)\geq f(a,z)+f(c,z).
    \]
    By symmetry, the same inequality holds when the variable occurs in the second position; a constraint using the variable in both positions also satisfies
    $2f(b,b)\geq f(a,a)+f(c,c)$.
    Fix any available assignment to the other variables, and let $C_x$ be its total cost when the variable in question receives $x$.
    Summing these inequalities with the constraint weights gives
    \[
        2C_b\geq C_a+C_c,
        \qquad\text{and hence}\qquad
        \min\{C_a,C_c\}\leq C_b.
    \]
    Thus $b$ can always be replaced by $a$ or $c$, even though neither value dominates $b$ and all pairs lie in a single reversing block.
\end{example}

We emphasize that this definition is a local property of a specific instance and not something that can be deduced from $\Gamma$ itself. For example, it is possible that an instance contains two variables $x$ and $y$, such that $a$ and $b$ are both available at $x$ and $y$, value $a$ is not locally necessary at $y$, while $b$ is not locally necessary at $x$.

\begin{example}\label{ex:variable_dependent_necessity}
    Let $f$ be the function from \cref{ex:local_necessity_single_block}, and let $h$ be obtained by interchanging $a$ and $b$ in both arguments:
    \[
        \begin{array}{c|rrr}
            h & a & b & c \\\hline
            a & 4 & 1 & 1 \\
            b & 1 & 0 & 2 \\
            c & 1 & 2 & 0
        \end{array}.
    \]
    Let $\Gamma$ consist of $f$, $h$, and all hard lists.
    The domain is still pairwise interacting and the move graph still has one reversing block, since the witnesses using $f$ remain available.
    Consider the instance on three variables $x,z,y$, each with available set $D=\{a,b,c\}$, whose total cost is
    \[
        f(x,z)+h(z,y).
    \]
    At $x$, the value $b$ is not locally necessary, by the inequality in the preceding example.
    However, $a$ and $c$ are locally necessary at $x$: they are uniquely cheapest when $z=a$ and $z=c$, respectively.
    At $y$, the roles of $a$ and $b$ are interchanged: $a$ is not locally necessary, whereas $b$ and $c$ are uniquely cheapest when $z=b$ and $z=c$, respectively.
    Thus the only unnecessary value at $x$ is $b$, while the only unnecessary value at $y$ is $a$; neither value is locally unnecessary at all variables of the instance.
    In fact, all three values are locally necessary at $z$, witnessed by $(x,y)=(b,b)$, $(a,a)$, and $(c,c)$ for $z=a,b,c$, respectively.
\end{example}

The significance of locally necessary values comes from the following crucial result. We show that if $I$ is an instance of \vcsp{}, $B$ is a reversing block, and every available set contains only pairwise interacting values belonging to $B$ (or has size one), and every available value is locally necessary, then every available set is expressive. Together with the fact that there is a reversing move on some $a,b$ belonging to $B$ and there is a move from any $\{c,d\}$ to $\{a,b\}$, it further follows that $\power{\Gamma}$ contains every finite-valued cost function on each available set.
\begin{theorem}[Follows from \cref{thm:fractional_necessity_expressive,obs:fractional_necessity_safe}]
    \label{thm:necessity_expressive}
    Let $D$ be a finite domain and $\Gamma$ a finite, finite-valued hard-list language over $D$, and $B$ a reversing block of $G_\Gamma$.
    Let $I$ be an instance of \vcsp{} such that the available set $L_v$ of each variable $v$:
    \begin{bracketenumerate}
        \item belongs to $B$ or has size one,
        \item is pairwise interacting, and
        \item has the property that each element of $L_v$ is locally necessary at $v$.
    \end{bracketenumerate}
    Then, every available set of $I$ is expressive.
\end{theorem}

This statement can be proved in an inductive way. The induction considers all nonempty subsets of the available sets; restricting a variable to a subset preserves local necessity for its remaining values. Consider a $k$-element available set $L_v=\{a_1,\dots,a_k\}$. As every $a_i$ is locally necessary, there is a partial assignment $\pi_i$ on the remaining variables such that assigning $a_i$ to $v$ is the unique minimum-cost extension of $\pi_i$ to the full instance $I$. To express any unary cost function on $L_v$, it is sufficient to express, for every $i$, a unary cost function $f_i$ such that $f_i(a_i)>f_i(a_j)$ for any $j\neq i$, and $f_i(a_j)=f_i(a_{j'})$ for any $j,j'\neq i$: nonnegative linear combinations of such $f_i$'s (with an additive shift) can express any unary function. We obtain such an $f_i$ as follows. Consider the cost function $g$ on the variables $V\setminus \{v\}$ which has some finite cost if the assignment on $V\setminus \{v\}$ is the same as some $\pi_{i'}$ for $i\neq i'$, and a prohibitively large cost otherwise. If we can add a constraint to the instance with such a cost function $g$, then this expresses a unary constraint on $v$. This unary constraint has the property that $a_i$ has strictly larger cost than $a_{i'}$ for any $i\neq i'$: no matter which of the $k-1$ reasonable assignments $\pi_{i'}$ is selected on $V\setminus \{v\}$, assigning $a_{i'}$ to $v$ is strictly better than assigning $a_i$. Furthermore, by modifying the value of $g$ when the assignment is equal to $\pi_{i'}$, it is not difficult to ensure that the resulting unary cost function on $v$ has the same value on $a_{i'}$ for every $i\neq i'$, and it is strictly larger on $a_i$.

Therefore, expressing all finite-valued unary cost functions on $\{a_1,\dots, a_k\}$ boils down to expressing such a cost function $g$. This seems to be a much harder task, as it is a cost function with high arity. However, we have made progress in an important sense. The cost $g$ needs to be low only on $k-1$ different assignments; in particular, for every variable $v'\in V\setminus \{v\}$, there is a list $L'_{v'}$ of size at most $k-1$ such that we can introduce a hard list restricting the domain of $v'$ to $L'_{v'}$. By induction, we may assume that we can express every finite-valued unary cost function on every $L'_{v'}$. Therefore, using the fact that every nonsingleton $L'_{v'}$ belongs to the reversing block $B$ and using arguments as in the soft-list case, we can show that we can express the cost function $g$. That is, expressing unary cost functions on a $k$-element set reduces to being able to express unary cost functions on every set of at most $k-1$ elements. For $k=2$, we can prove this directly using the definition of reversing blocks, establishing the base case of the induction.
\paragraph{Locally fractionally necessary values}
This discussion suggests the following algorithm: after solving the nonreversing-block subinstances in polynomial time, let us repeatedly remove any value that is not locally necessary in the remaining subinstances, and then solve the problem using the standard dynamic programming approach.
As every value is locally necessary, every available set is expressive, hence its size is not larger than $c_{\Gamma}$ (recall \cref{thm:hard_lists_base_equality}) resulting in the required running time. However, it is not clear how to efficiently check if a value $a$ is locally necessary at variable $v$: testing all possible partial assignments as in the definition is clearly not feasible. In fact, this is an \np-hard problem in general.

Instead of exactly determining if a value is locally necessary, we test for a weaker condition. Testing whether $a$ is locally necessary at $v$ can be formulated as testing whether the optimum of a mixed integer linear program is positive (see~\cref{milp:local_necessity}). We say that $a$ is \emph{locally fractionally necessary} at $v$ if the optimum of the LP relaxation of this mixed integer linear program is positive. This definition requires solving a linear program, so the condition can be tested in polynomial time. Clearly, local necessity implies local fractional necessity. Therefore, we can safely remove values from the available set if they are not even locally fractionally necessary, giving us a new polynomial-time preprocessing step. The test is repeated after each deletion until no further value can be removed (\cref{thm:remove_fractionally_unnecessary}).

With this preprocessing, we achieve the weaker property that every value in every available set is locally fractionally necessary. However, the proof that every available set is expressive can be made to work under this weaker condition: \cref{thm:necessity_expressive} remains true even if we weaken \emph{locally necessary} to \emph{locally fractionally necessary}
(see \cref{thm:fractional_necessity_expressive}). Intuitively, a fractional solution of the LP gives an assignment that does not give a single value to each variable of $V\setminus \{v\}$, but instead assigns distributions over tuples to the incident constraints. Using that the fractional solution is rational, we can duplicate variables and constraints appropriately to obtain an instance $I'$ and
an assignment showing that the value is locally necessary at $v$ in $I'$. We can use such blown-up instances in the proof showing that every finite-valued unary cost function on the available set is expressible.

Therefore, it is sufficient to ensure that every value in every available set is locally fractionally necessary.
Every available set is then expressive, and hence has size at most $c_{\Gamma}$.
Therefore, after this preprocessing, the standard dynamic programming approach proves the algorithmic part of \cref{thm:main}.

\section{Preliminaries}
\label{sec:preliminaries}
\paragraph*{VCSPs.}
A VCSP is the natural optimization
variant of a CSP~\cite{cohenComplexitySoftConstraint2006}.
A cost function is a function $f: D^{a} \to \ratNnonneg \cup \{\infty\}$, where $a \geq 0$ is the \emph{arity} of $f$, also denoted by $\arity{f}$.
For an integer $n\geq 0$, write $[n]=\{1,\dots,n\}$, so $[0]=\varnothing$.
For a fixed finite domain $D$ and a finite set of cost functions $\Gamma$, an instance $I=(V,\CC,t)$ of \vcsp{} consists of a finite set of variables $V$, a finite family of constraints $\CC$, and a cost target $t \in \ratN$.
Each constraint $C \in \CC$ is a triple $(f,S,w)$, where $f \in \Gamma$ is a cost function denoted by $\costfun{C}$, $S \in V^{\arity{f}}$ is a tuple of $\arity{f}$ variables called the \emph{scope} of $C$, denoted by $\scope{C}$, and $w \in \ratNpositive$ is a positive rational weight, denoted by $\weight{C}$.
For a variable assignment $\pi:V\to D$, write $\cost_I(\pi)=\sum_{C\in\CC}\weight{C}\costfun{C}(\pi(\scope{C}))$, where $\pi$ acts componentwise on tuples.
The optimum cost is $\Opt(I)=\min_{\pi:V\to D}\cost_I(\pi)$, taking the value $\infty$ if no finite-cost assignment exists.
We also use $\cost$ and $\Opt$ for a weighted constraint system $(V,\CC)$ without a target, using the same formulas.
The decision problem asks whether $\Opt(I)\leq t$.
We write $|I|$ for the bit length of the encoding of $I$.
The \emph{primal graph} of an instance has vertex set $V$ and connects two
distinct vertices with an edge if they appear together in a constraint scope.
Since all costs are nonnegative, every instance with a negative target is a no-instance.
Throughout the paper, we shall assume all domains are nonempty and finite.

\begin{remark}
    Requiring positive rational weights for the constraints is equivalent to requiring
    positive integer weights.
    Indeed, multiplying all weights and the target
    by a common denominator of the weights yields an equivalent instance.
    We allow rational weights for convenience, as they simplify the proofs.

    Similarly, the assumption that the costs are nonnegative is without loss of generality; if negative costs are involved they can be cleared by appropriate shifting.
\end{remark}

A cost function is \emph{finite-valued} if it never takes the value $\infty$.
A constraint language $\Gamma$ is \emph{finite-valued} if all its cost functions are finite-valued.
For a cost function $f: D^r \to \ratNnonneg \cup \{\infty\}$ we define $\max f = \max_{\bar x \in D^r} f(\bar x)$, similarly for $\min f$.

We write $\power{\Gamma}$ for the expressive power~\cite{CCCJZ13,thapperComplexityFiniteValued2016} of $\Gamma$: cost functions obtained from finite weighted sums of constraints from $\Gamma$ by minimizing over auxiliary variables, allowing an additive rational shift provided that the resulting cost function is nonnegative.
Hence, for each $g \in \power{\Gamma}$ of arity $d$ there exists some gadget with distinguished ``input'' variables $x_1,\dots,x_d$, and auxiliary variables over which the cost is optimized, such that for any assignment $\bar y$ to $x_1,\dots,x_d$ the corresponding cost is $g(\bar y)$, up to an additive constant.
The following simple lemma allows us to use a fixed \emph{finite} family of expressed cost functions in reductions.

\begin{lemma}
    \label{thm:expressive_power_reduction}
    Let $D$ be a fixed finite domain, $\Gamma$ a fixed constraint language over $D$, and $\Delta\subseteq\power{\Gamma}$ a fixed finite set of cost functions.
    There is a polynomial-time algorithm that, given an instance $I$ of \vcsp[\Gamma\cup\Delta]{} and a path or tree decomposition of its primal graph of width $k$, constructs an equivalent instance of \vcsp[\Gamma]{} together with a decomposition of the same type and width at most $k+\Oh(1)$.
    The constant in the width bound depends only on $\Delta$.
\end{lemma}
\begin{proof}
    The statement is trivial if $\Delta = \emptyset$, so we assume $\Delta$ is nonempty.
    For each $f\in\Delta$, fix a gadget over $\Gamma$ with cost $F_f(\bar x,\bar y)$ and a rational constant $c_f$ such that
    \[
        f(\bar x)=\min_{\bar y}F_f(\bar x,\bar y)+c_f,
    \]
    where $\bar x$ are its input variables and $\bar y$ are its auxiliary variables.
    Replace every constraint $(f,S,w)$ with $f\in\Delta$ by a separate copy of this gadget, identifying its inputs with $S$, using fresh auxiliary variables, and multiplying every gadget weight by $w$.
    Keep all other constraints and set the new target to
    \[
        t'=t-\sum_{\substack{(f,S,w)\in\CC,\\f\in\Delta}}w c_f,
    \]
    where $\CC$ is the constraint family of $I$ and $t$ its target.
    For every assignment to the original variables, the separate auxiliary variables can be minimized independently.
    The minimum resulting cost is the original cost minus the same sum of weighted constants.
    This shows that the two instances are equivalent.

    Let $K$ be the largest number of auxiliary variables in one of the fixed gadgets.
    For each replaced constraint, its distinct variables form a clique in the original primal graph, so choose a bag containing the constraint scope.
    For a nullary constraint, use any bag (creating an empty bag if needed).

    In a path decomposition, insert a copy of the chosen bag immediately after it and add all auxiliary variables of the gadget added for the constraint to the copy.
    Use a separate bag copy for each gadget.
    This preserves the property that each original variable appears in the bags of a connected subpath of the path decomposition, and every auxiliary variable appears in just its gadget's bag.

    In a tree decomposition, attach a new leaf bag containing the gadget's inputs and auxiliary variables to the chosen bag.
    The bags containing each original variable remain connected, and each auxiliary variable appears only in its leaf.
    In both cases, the new bags cover all gadget constraints and have size at most $k+1+K$.
    Thus the width increases by at most the constant $K$.

    Since $\Delta$ is fixed and finite, every replaced constraint produces only constantly many constraints and auxiliary variables.
    Hence, the reduction takes polynomial time.
\end{proof}

\paragraph*{Lists.}
Next, we formally define the notions of hard and soft lists, and the notion of a finite-valued hard-list language.

\begin{restatable}[Hard lists]{definition}{defHardLists}
    \label{def:hard_lists}
    Let $D$ be a finite domain.
    A hard list on $D' \subseteq D$ is the cost function $f$ defined by $f(d) = 0$ if $d \in D'$ and $f(d) = \infty$ if $d \notin D'$.
    A constraint language $\Gamma$ over $D$ contains all hard lists if, for all $D'
        \subseteq D$, the language $\Gamma$ contains the hard list on $D'$.
\end{restatable}

\begin{restatable}[Finite-valued hard-list language]{definition}{defHardListLanguage}
    \label{def:hard_list_language}
    Let $D$ be a finite domain and let $\mathcal H_D$ be the set of all hard lists on $D$.
    A constraint language $\Gamma$ over $D$ is a \emph{finite-valued hard-list language} if $\Gamma=\Phi\cup\mathcal H_D$ for some finite-valued constraint language $\Phi$ over $D$.
    Equivalently, $\Gamma$ contains all hard lists and every cost function in $\Gamma\setminus\mathcal H_D$ is finite-valued.
    In an instance over $\Gamma$, a \emph{non-hard-list constraint} is a constraint whose cost function belongs to $\Gamma\setminus\mathcal H_D$.
\end{restatable}

Since hard lists can be seen as simply forbidding certain variable assignments, this leads to a natural notion of available elements.

\begin{restatable}[Available elements]{definition}{availableElements}
    \label{def:available_elements}
    Let $D$ be a finite domain, let $\Gamma$ be a constraint language over $D$ whose cost functions are finite-valued or hard lists, and let $I$ be an instance of \vcsp{}.
    The \emph{available set} $L_v$ of a variable $v$ consists of those elements $d\in D$ for which assigning $d$ to $v$ does not by itself make any constraint involving $v$ incur infinite cost.
    Equivalently, $L_v$ is the intersection of the sets allowed by the hard-list constraints imposed on $v$, with $L_v=D$ if there are no such constraints.
    An element $d$ is \emph{available at $v$} if $d\in L_v$.
\end{restatable}

Next, we also formally define soft lists and soft-list languages.

\begin{restatable}[Soft lists and soft-list languages]{definition}{defSoftLists}
    \label{def:soft_lists}
    Let $B\subseteq D$.
    The \emph{soft list on $B$} is the unary cost function
    \[
        h_B(d)=
        \begin{cases}
            0, & d\in B,    \\
            1, & d\notin B.
        \end{cases}
    \]
    We say that $\Gamma$ \emph{expresses all soft lists} if $h_B\in\power{\Gamma}$ for every $B\subseteq D$.
    If $h_B\in\Gamma$ for every $B\subseteq D$, we say that $\Gamma$ \emph{contains all soft lists}.

    For $B\subseteq D$, we say that $\Gamma$ \emph{expresses all soft lists on $B$} if, for every $S\subseteq B$, there is a finite-valued unary function $g\in\power{\Gamma}$ and a constant $c\in\ratN$ such that $g(d)=h_S(d)+c$ for every $d\in B$.

    We call a finite constraint language $\Gamma$ over $D$ a \emph{soft-list language} if, for every $d\in D$, $\Gamma$ contains the unary cost function that takes value $1$ at $d$ and $0$ at every other element of $D$.
\end{restatable}

\paragraph*{Treewidth and pathwidth.}
For an introduction to treewidth and pathwidth, see, e.g.,~\cite[Section 7]{CyganFKLMPPS15}.
For a supplied tree or path decomposition $\mathcal T$, we write $|\mathcal T|$ for the bit length of its encoding, including its bags and underlying tree or path.
We show lower bounds based on \pwseth{} and \twseth{}, which we state formally below.
The \pwseth{} was introduced by Lampis~\cite{lampisPrimalPathwidthSETH}; the \twseth{} was postulated in the work of Iwata and Yoshida~\cite{IwataY15TWSETH}, and recently explored further by Lampis~\cite{lampisTreewidthSETH}.

\begin{hypothesis}[\pwseth{}]
    \label{hypothesis:pwseth}
    For all $\eps > 0$ there is no algorithm that can solve \threeSat{} in time
    \[
        (2-\eps)^{\pw}\cdot (|I|+|\mathcal T|)^{\Oh(1)},
    \]
    where the input consists of a \threeSat{} instance $I$ and a path decomposition $\mathcal T$ of its primal graph of \mbox{width $\pw$}.
\end{hypothesis}

\begin{hypothesis}[\twseth{}]
    \label{hypothesis:twseth}
    For all $\eps > 0$ there is no algorithm that can solve \threeSat{} in time
    \[
        (2-\eps)^{\tw}\cdot (|I|+|\mathcal T|)^{\Oh(1)},
    \]
    where the input consists of a \threeSat{} instance $I$ and a tree decomposition $\mathcal T$ of its primal graph of \mbox{width $\tw$}.
\end{hypothesis}

Lampis~\cite{lampisPrimalPathwidthSETH} showed that the \pwseth{} is implied by the SETH (whose refutation is reported in recent preprints~\cite{OpenAI26ETH,ClaudeAnthropic26ETH}), the $k$-Orthogonal Vectors Assumption~\cite{VassilevskaWilliams18}, and the Set Cover Conjecture~\cite{CyganEtAl16}.
The \twseth{} is implied by the \pwseth{}, and Lampis~\cite{lampisTreewidthSETH} showed that it is also implied by the \textsc{$k$-Sum} Hypothesis~\cite{AbboudLewi13} and the \textsc{$k$-Xor} Hypothesis~\cite{DalirrooyfardLincolnSahaWilliams25}.
The traditional variants of the \textsc{$k$-Sum} and \textsc{$k$-Xor} hypotheses were recently proven false, because they are wrong when $k = 3$~\cite{alman2026trulysubquadratic3sumtruly}.
A recent preprint~\cite{chan20263sumsimplerwithoutmatrix} also disproves the \textsc{$k$-Sum} Hypothesis for all odd $k$.
However, as explicitly discussed in~\cite{lampis2026ksumhardnessimpliestreewidthsetharxiv}, the updated arXiv version of~\cite{lampisTreewidthSETH}, for the result of Lampis the following weaker variants suffice: for each $k \geq 3$ the problem \textsc{$k$-Sum}, respectively \textsc{$k$-Xor}, cannot be solved in time $O(n^{k/2 - \varepsilon})$, where $n$ is the number of integers/vectors in each of the $k$ given input lists, for any $\varepsilon > 0$.
These variants are, as of now,  not refuted by these results.
We use both the \twseth{} and \pwseth{} because they offer different advantages: \pwseth{} yields lower bounds in terms of pathwidth, whereas \twseth{} yields treewidth lower bounds under a weaker assumption.
As discussed in \cref{sec:intro}, \twseth{} and \pwseth{} are also implied by CNF-SETH, which has not been disproved in the recent results~\cite{OpenAI26ETH,ClaudeAnthropic26ETH}.
\section{Interaction and Moves}
\label{sec:interaction_moves}

Before proving the results for finite-valued soft-list languages and finite-valued hard-list languages, we establish the expressibility results needed in both settings.
First, we introduce the notion of a restriction to capture both soft and hard lists.

\begin{definition}[Restrictions]
    \label{def:restriction}
    Let $B\subseteq D$ be nonempty.
    A \emph{restriction to $B$} is a unary cost function $h$ for which there is a constant $c\in\ratN$ such that $h(d)=c$ for $d\in B$ and $h(d)>c$ for $d\notin B$; costs outside $B$ can be infinite.
    We say that $\Gamma$ expresses a restriction to $B$ if such a function belongs to $\power{\Gamma}$.
\end{definition}

Observe that every soft or hard list on $B$ is a restriction to $B$.
Our results revolve around being able to express all cost functions on certain domain subsets, which we formally define next.

\begin{definition}[Cost functions on a subset]
    \label{def:costs_subset}
    \label{def:unary_costs_subset}
    Let $B\subseteq D$.
    We say that $\Gamma$ \emph{expresses all finite-valued cost functions on $B$} if, for every $r\geq 1$ and every function $f:B^r\to\ratNnonneg$, there are a finite-valued cost function $g\in\power{\Gamma}$ and a constant $c\in\ratN$ such that $g(\bar d)=f(\bar d)+c$ for all $\bar d\in B^r$.
    If we require this only for $r=1$, we say that $\Gamma$ \emph{expresses all finite-valued unary cost functions on $B$}.
\end{definition}

Expressing all finite-valued unary cost functions on $B$ lets us prescribe costs within $B$, without any requirement on the costs outside $B$.
Hence, it is quite different from restrictions, which restrict the optimum assignments and concern the entire domain.
Next, we introduce the notion of a move, following the work of Esmer, Focke, Marx, and Rzążewski~\cite{esmerListHomomorphismsDeleting2024ESA}
and
the idea, sometimes called a max-cut condition, from the work of Kolmogorov and \v{Z}ivn{\'y}~\cite{KZ13}, Huber, Krokhin, and Powell~\cite{HKP14:sicomp}, and Thapper and \v{Z}ivn{\'y}~\cite{thapperComplexityFiniteValued2016}.

\movesAndReversingMoves*

The defining condition of a move is unchanged when $a$ and $b$ are swapped
and, simultaneously, $c$ and $d$ are swapped.
Hence a move from $(a,b)$ to
$(c,d)$ is also a move from $(b,a)$ to $(d,c)$, and
$\{a,b\}\moves \{c,d\}$ holds if and only if there is a move to
$(c,d)$ from $(a,b)$ or from $(b,a)$.

Now, we establish some further basic properties of moves, concretely that they
are symmetric and compose,
cf.~\cite[Lemma~5.1(b)]{KZ13}.

\begin{lemma}
    \label{thm:move_composition}
    Let $D$ be a finite domain and $\Gamma$ a constraint language over $D$.
    If $(a,b) \moves (c,d)$, then $(c,d) \moves (a,b)$.
    Moreover, if there are finite-valued moves from $(a,b)$ to $(c,d)$ and from $(c,d)$ to $(x,y)$ in $\power{\Gamma}$, and $\Gamma$ expresses a restriction to $\{c,d\}$, then there is a finite-valued move from $(a,b)$ to $(x,y)$ in $\power{\Gamma}$.
\end{lemma}
\begin{proof}
    Swapping the two arguments of a move from $(a,b)$ to $(c,d)$ gives a move
    from $(c,d)$ to $(a,b)$.
    For composition, let $f,g\in\power{\Gamma}$ be the two
    moves, with $f(a,c)=f(b,d)=\alpha<\alpha'=f(a,d)=f(b,c)$ and
    $g(c,x)=g(d,y)=\beta<\beta'=g(c,y)=g(d,x)$.
    Let $h\in\power{\Gamma}$ be a
    restriction to $\{c,d\}$, shifted to be zero on $\{c,d\}$, and choose a
    rational $\lambda>0$ with $\lambda h(z)>\max f+\max g$ for all
    $z\notin\{c,d\}$.
    Then
    \[
        k(s,t)=\min_{z\in D}\bigl(f(s,z)+g(z,t)+\lambda h(z)\bigr)
    \]
    is a finite-valued function in $\power{\Gamma}$, and its minimum is always
    attained at some $z\in\{c,d\}$.
    Hence
    $k(a,x)=k(b,y)=\alpha+\beta<\min\{\alpha+\beta',\alpha'+\beta\}=k(a,y)=k(b,x)$,
    so $k$ is a move from $(a,b)$ to $(x,y)$.
\end{proof}

Recall that our approach relies on a sequence of preprocessing steps.
The first one concerns the domination of elements, defined as follows.

\begin{definition}[Domination]
    \label{def:domination_shared}
    Let $D$ be a finite domain and $\Gamma$ a constraint language over $D$.
    An element $a\in D$ \emph{dominates} an element $b\in D$ if $h(a)\leq h(b)$ for every finite-valued unary cost function $h\in\power{\Gamma}$.
\end{definition}

For finite-valued hard-list languages, a dominated available element can be deleted when a distinct dominating element is also available at the same variable (\cref{thm:remove_dominated_elements}).
Crucially, two domain elements that do not dominate each other allow us to express all finite-valued unary cost functions on their pair.

\begin{lemma}
    \label{thm:undominated_pair_unaries}
    Let $D$ be a finite domain and $\Gamma$ a constraint language over $D$.
    If distinct $a,b\in D$ do not dominate each other, then $\Gamma$ expresses all finite-valued unary cost functions on $\{a,b\}$.
\end{lemma}
\begin{proof}
    By \cref{def:domination_shared}, there are finite-valued unary functions $h_a,h_b\in\power{\Gamma}$ with $h_a(a) < h_a(b)$ and $h_b(b)<h_b(a)$.
    Let $u:\{a,b\}\to\ratNnonneg$ be an arbitrary function, and put $\delta=u(b)-u(a)$.
    If $\delta>0$, scale $h_a$ by $\delta/(h_a(b)-h_a(a))$; if $\delta<0$, scale $h_b$ by $\delta/(h_b(b)-h_b(a))$; and if $\delta=0$, use the constant zero function.
    In every case the resulting finite-valued $g\in\power{\Gamma}$ satisfies $g(b)-g(a)=\delta$.
    Taking $c=g(a)-u(a)$ gives $g(x)=u(x)+c$ on $x \in \{a,b\}$.
\end{proof}

The next preprocessing step we use concerns the separability of cost functions.
For the definitions here, it suffices to focus on the binary case.
Consider some cost function $f$ of the constraint language.
In principle, having a constraint in which $f$ acts on two variables $x$ and $y$ couples these variables (which is expressed by an edge in the primal graph).
However, if $f(x,y) = u(x) + v(y)$ for two expressible unary cost functions $u$, $v$, then in an instance we can replace occurrences of $f$ with two unary functions.
If this is the case for all cost functions acting on both $x$ and $y$, then we can preprocess the instance so that the remaining primal graph does not have an edge from $x$ to $y$.
Such functions $f$ are also called separable~\cite{Topkis98} or modular~\cite{BorosH02,HelaouiNaanaa13}.
More specifically, for a finite-valued binary cost function $f$, Helaoui and Naanaa call $a,b$ a \emph{modular pair} if $f$ is modular on $\{a,b\}\times D$~\cite{HelaouiNaanaa13}.
They show that this relation is an equivalence relation and exploit modular classes in search algorithms for binary VCSPs.
We use this viewpoint to define interaction between domain elements and additionally impose non-domination conditions.

\interactingPairs*

In our algorithm, we will decompose the instances based on noninteraction.
For now, we focus on the opposite task: establishing properties we obtain based on interaction.
First, we show that interaction gives a move to its witness pair, which explains the importance of this notion.

\begin{lemma}
    \label{thm:interaction_move}
    Let $D$ be a finite domain and $\Gamma$ a constraint language over $D$.
    Let $a,b\in D$ have interaction witnessed by $c,d$ and a finite-valued binary function $f\in\power{\Gamma}$.
    If $f(a,c)-f(b,c)<f(a,d)-f(b,d)$, there is a finite-valued move $(a,b) \moves (c,d)$.
    Otherwise, there is a finite-valued move $(a,b) \moves (d,c)$.
    In any case, there is a finite-valued move $\{a,b\} \moves \{c,d\}$.
\end{lemma}
\begin{proof}
    For unary functions $u,v$, the function $g(x,y)=f(x,y)+u(x)+v(y)$
    satisfies
    \[
        g(a,c)+g(b,d)-g(a,d)-g(b,c)=f(a,c)+f(b,d)-f(a,d)-f(b,c)=:\Delta,
    \]
    and $\Delta\ne0$ by the witness assumption.
    Since neither element of
    $\{a,b\}$, nor of $\{c,d\}$, dominates the other,
    \cref{thm:undominated_pair_unaries} gives finite-valued unary $u,v\in\power{\Gamma}$
    with
    \[
        \begin{aligned}
            u(b)-u(a) & =\tfrac12\bigl(f(a,c)-f(b,d)+f(a,d)-f(b,c)\bigr), \\
            v(d)-v(c) & =\tfrac12\bigl(f(a,c)-f(b,d)-f(a,d)+f(b,c)\bigr).
        \end{aligned}
    \]
    This choice makes $g(a,c)=g(b,d)$ and $g(a,d)=g(b,c)$, so
    $g(a,c)-g(a,d)=\Delta/2$.
    Hence the finite-valued function~$g$ is a move
    from $(a,b)$ to $(c,d)$ if $\Delta<0$, that is, if
    $f(a,c)-f(b,c)<f(a,d)-f(b,d)$, and a move from $(a,b)$ to $(d,c)$
    otherwise.
    Similar normalizations appear in the proofs of~\cite[Lemma~5.1(b)]{KZ13} and~\cite[Lemma~2.4]{thapperComplexityFiniteValued2016}.
\end{proof}

Having established the required moves based on interaction, we turn to expressing all finite-valued cost functions, first on a two-element set and then on larger sets.

\begin{lemma}
    \label{thm:boolean_costs}
    Let $D$ be a finite domain, $\Gamma$ a constraint language over $D$, and $P=\{p,q\}\subseteq D$ with $p\neq q$.
    Suppose that $\Gamma$ expresses all finite-valued unary cost functions on $P$, a finite-valued reversing move on $P$, and a restriction to $P$.
    Then $\Gamma$ expresses all finite-valued cost functions on $P$.
\end{lemma}
\begin{proof}
    Identify $p,q$ with $0,1$, and let $h\in\power{\Gamma}$ be a restriction
    to~$P$, shifted to be zero on~$P$.
    Every auxiliary variable below is
    forced into~$P$ by adding $\lambda h$ for a sufficiently large
    rational~$\lambda$.
    All other terms are finite-valued, so this keeps the
    expressed functions finite-valued, and for inputs in~$P$ they agree with the
    displayed expressions up to additive constants.

    Rescaling the reversing move gives $e\in\power{\Gamma}$ with $e(s,t)=1$
    for $s=t$ and $e(s,t)=0$ for $s\ne t$, on~$P^2$ and up to a constant.
    Its
    complement is $1-e(s,t)=\min_{z\in P}\bigl(2e(s,z)+e(z,t)\bigr)$, since
    the minimum is attained at $z\ne s$.
    On~$P^2$ we have
    \[
        st=\tfrac12\bigl(e(s,t)+s+t-1\bigr),\qquad
        -st=\tfrac12\bigl(1-e(s,t)-s-t\bigr).
    \]
    Together with the unary functions on~$P$, these give every polynomial of
    degree at most two in variables ranging over~$P$, up to a constant.
    In particular, $\Gamma$ expresses all finite-valued unary and binary cost
    functions on~$P$.

    Let $r\ge1$ and $f\colon P^r\to\ratNnonneg$.
    By~\cite[Theorem~28]{cohenExpressivePowerValued2008}, $f$ can be expressed
    using finite-valued unary and binary cost functions on~$P$.
    Replace each cost function in such a gadget by its finite-valued
    representative in $\power{\Gamma}$, and restrict its auxiliary variables
    to~$P$ as above.
    The resulting function belongs to $\power{\Gamma}$ and
    agrees with $f$ up to a constant on~$P^r$.
    It is finite-valued on all of
    $D^r$, since every input assignment admits a finite-cost extension with
    all auxiliary variables in~$P$.
\end{proof}

Next, we extend this construction to larger sets.
We follow the indicator strategy of~\cite{esmerListHomomorphismsDeleting2024ESA}: combine moves that distinguish pairs of inputs into an encoding, and use unary penalties to make the minimum cost independent of the input.
In their setting, these penalties come from pendant vertices with suitable lists; here they are supplied by the assumed expressibility of unary cost functions.

\begin{lemma}
    \label{thm:transfer_costs}
    Let $D$ be a finite domain, $\Gamma$ a constraint language over $D$, and $P=\{p,q\}\subseteq D$ with $p\neq q$.
    Let $r\geq 1$ and let $S_1,\dots,S_r\subseteq D$ be nonempty subsets.
    Suppose that $\Gamma$ expresses all finite-valued cost functions on $P$ and all finite-valued unary cost functions on each $S_i$.
    Assume that $\Gamma$ expresses a restriction to $P$.
    Suppose further that there is a finite-valued move $\{a,b\}\moves P$ for every two distinct elements $a,b\in S_i$, for each $i$.
    For every $f:S_1\times\cdots\times S_r\to\ratNnonneg$, there are a finite-valued $g\in\power{\Gamma}$ and a constant $c\in\ratN$ with $g(\bar x)=f(\bar x)+c$ on $\bar x \in S_1\times\cdots\times S_r$.
\end{lemma}
\begin{proof}
    Coordinates $i \in [r]$ with $|S_i| = 1$ can be substituted in $f$ and then ignored in the gadget.
    If every $S_i$ has size one, $f$ is constant on $S_1 \times \cdots \times S_r$, and the zero-cost gadget suffices.
    We may therefore assume that each $S_i$ has at least two elements.

    For each $i \in [r]$ and each pair of distinct elements $a,b\in S_i$, choose an ordering $(a,b)$ and a finite-valued move $m_{(a,b) \moves (p,q)}\in\power{\Gamma}$ from $(a,b)$ to $(p,q)$.
    Let $\mathcal{S}_i$ be the set of the chosen ordered pairs of $S_i$, and put $k_i=|\mathcal{S}_i|=\binom{|S_i|}{2}$.
    For each $i$, we use a common input variable $x_i$ and a separate output variable $z_{i,(a,b)}$ for each $(a,b)\in\mathcal{S}_i$.
    Fix an ordering of $\mathcal{S}_i$ and write $\bar z_i=(z_{i,(a,b)})_{(a,b)\in\mathcal{S}_i}$.
    Define the finite-valued cost function $H_i\in\power{\Gamma}$ by
    \[
        H_i(x_i,\bar z_i)=\sum_{(a,b)\in\mathcal{S}_i}
        m_{(a,b)\moves(p,q)}(x_i,z_{i,(a,b)}).
    \]
    That is, we sum up over all moves from $\mathcal{S}_i$ to $P$ and identify the first variable while keeping the second variable of each move separate.
    Intuitively, the first variable $x_i$ is the input of $H_i$ and the variables $\bar z_i$ are the output variables.
    For each pair $\{a,b\}$ of elements of $S_i$ the function has one dedicated output variable.

    We will force every output variable to lie in $P$ once the rest of the gadget is constructed.
    Write
    \[
        M_i(a)=\min_{\bar z\in P^{k_i}}H_i(a,\bar z),
    \]
    for the minimum cost obtainable when assigning domain element $a$ to the input of the construction for $i$.
    For each $a\in S_i$ select a minimizing tuple $\bar z_i(a) \in P^{k_i}$, i.e., a tuple $\bar z_i(a)$ such that $H_i(a,\bar z_i(a)) = M_i(a)$.
    The output variable $z_{i,(a,b)}$ has different unique minimizing values for inputs $a$ and $b$ due to the move $m_{(a,b) \moves (p,q)}$ acting on it.
    Since all the moves used in the construction have separate outputs, their minima are attained independently.
    Consequently, there is no assignment $\bar z \in P^{k_i}$ to the outputs that minimizes $H_i$ for two distinct inputs $a,b \in S_i$.
    Hence,
    \[
        \delta_i=
        \min_{\substack{a,b\in S_i\\a\neq b}}
        \bigl(H_i(a,\bar z_i(b))-M_i(a)\bigr)>0.
    \]
    In particular, the selected tuples $\bar z_i(a)$ are all distinct.

    So far, we ensured that for any two distinct assignments to the input variables that come from the domain $S_1 \times \cdots \times S_r$, the sets of assignments to the output variables, restricted to assignments using domain elements of $P$, that minimize the cost of the constructed gadget do not intersect.
    However, it may still be the case that certain inputs are cheaper overall than others.

    To mitigate this, use the assumed unary expressibility on each $S_i$ to add, up to constants, the nonnegative unary costs
    \[
        \max_{a\in S_i}M_i(a)-M_i(x_i)
        \qquad\text{on }S_i.
    \]
    These normalize the input-dependent minima, guaranteeing that all possible assignments from $S_i$ to input variable $x_i$ have the same cost $C_i$ when using intended assignments to the outputs.
    Call the gadget created for $S_i$ so far the encoding gadget for $S_i$.

    Put $R=\max f$ and choose positive rational weights $\lambda_i$ with $\lambda_i\delta_i>R$.
    We weigh each used encoding gadget with $\lambda_i$.
    This will ensure that, for input assignment $a \in S_i$, it comes with a significant cost to assign a tuple $\bar z_i(b)$ for $b \in S_i \setminus \{a\}$ to the outputs.

    Next, we will use the expressibility on $P$ to ensure the overall gadget expresses the function $f$ on the intended domain.
    By this assumed expressibility, express on the concatenated output tuples $\bar z_1,\dots, \bar z_r$ the function that has value
    \[
        f(a_1,\dots,a_r)
        \quad\text{at }(\bar z_1(a_1),\dots,\bar z_r(a_r))
    \]
    and value $R+1$ at every other tuple in $P^{k_1+\cdots+k_r}$.

    First consider minimizing the resulting gadget with all outputs taking values from $P$.
    For inputs $x_i\in S_i$, their selected tuples $\bar z_i(x_i)$ attain, up to a common constant, the value
    \[
        \sum_i\lambda_i C_i+f(x_1,\dots,x_r),
    \]
    where the cost $\sum_i \lambda_i C_i$ stems from the encoding gadgets and the cost of $f$ from the function we added in the previous step.
    Any selected tuple belonging to a different input (i.e., for some $i \in [r]$ the assignment to the output variables $\bar z_i$ is $\bar z_i(b)$ for some $b \in S_i$ different from the assignment to $x_i$) costs more than $\sum_i\lambda_i C_i+R$, because at least one of the encoding gadgets incurs a gap $\lambda_i\delta_i>R$.
    Every tuple not selected for any input (i.e., for some $i \in [r]$ the assignment to the output variables $\bar z_i$ is not of the form $\bar z_i(b)$ for any $b \in S_i$) pays the penalty $R+1$ in addition to the cost due to the encoding gadgets, which is itself at least $\sum_i \lambda_i C_i$.
    Hence neither kind of tuple can improve on the variable assignment intended for the actual input.

    Finally, we need to ensure that all of the output variables take a value from $P$.
    Let $h\in\power{\Gamma}$ be a restriction to $P$ which exists by assumption, shifted so that it is zero on $P$.
    Let $F$ be the cost function of the gadget constructed so far.
    Choose a positive rational $W$ such that $Wh(d)>\max F$ for every $d\notin P$.
    Add $Wh$ at every output variable.
    As in the proof of \cref{thm:boolean_costs}, this makes the minimum attained with all outputs in $P$.
    The restrictions act only on auxiliary variables, and every assignment to the exposed inputs has a finite extension with all outputs in $P$.
    So, the resulting function is finite-valued on $D^r$.
\end{proof}

\section{VCSP over Finite-Valued Soft-List Languages}
\label{sec:vcsp_soft_lists}
We now apply the results on interaction and moves to finite-valued soft-list languages.
Let us first recall the central definitions surrounding soft lists.

\defSoftLists*{}

Every finite-valued soft-list language expresses all soft lists: $h_B$ is the sum, over $d\in D\setminus B$, of the unary functions taking value $1$ at $d$ and $0$ elsewhere.

We will show that if two elements of the domain are equivalent in some sense, then we can preprocess one of them away.
Once we are left with an instance in which no two remaining elements are equivalent, we show a matching lower bound.
Intuitively, two elements $d_1$ and $d_2$ are equivalent if replacing $d_1$ with $d_2$ in any constraint changes the cost by a fixed constant that depends only on the ordered pair $d_1,d_2$, the cost function, and the position of the element in the constraint.

\begin{definition}[Additive equivalence]
    \label{def:additive_equivalence}
    Two elements $d_1,d_2 \in D$ are \emph{additively equivalent} if, for each $f \in \Gamma$ and each $i \in \{1,\dots,\arity{f}\}$, there is a constant $c^{d_1,d_2}_{f,i} \in \ratN$ such that, for all $w_1,\dots,w_{\arity{f}-1} \in D$, we have
    \[
        f(w_1,\dots,w_{i-1},d_1,w_i,\dots,w_{\arity{f}-1})
        = f(w_1,\dots,w_{i-1},d_2,w_i,\dots,w_{\arity{f}-1}) + c^{d_1,d_2}_{f,i}.
    \]
    If $d_1$ and $d_2$ are additively equivalent we also write $d_1 \additiveequiv d_2$.
    We call $c^{d_1,d_2}_{f,i}$ a constant witnessing $d_1\additiveequiv d_2$ for $f$ and position $i$.
\end{definition}

Clearly, additive equivalence defines an equivalence relation.

\begin{lemma}
    Let $D$ be a domain and $\Gamma$ a constraint language over $D$.
    Then, $\additiveequiv$ is an equivalence relation.
\end{lemma}
\begin{proof}
    The constants
    $c^{d,d}_{f,i}=0$, $c^{d_2,d_1}_{f,i}=-c^{d_1,d_2}_{f,i}$, and
    $c^{d_1,d_3}_{f,i} = c^{d_1,d_2}_{f,i} + c^{d_2,d_3}_{f,i}$ witness reflexivity,
    symmetry, and transitivity, respectively.
\end{proof}

\subsection{The Algorithm}

Now, let us work towards obtaining our algorithm.
First, we can show that if two distinct elements are additively equivalent, we can delete one of them without changing the problem.

\begin{lemma}
    \label{thm:add_equiv_reduction}
    Let $D$ be a fixed finite domain and $\Gamma$ a finite-valued soft-list language over $D$.
    Suppose that two distinct elements $d_1,d_2\in D$ are additively equivalent.
    Then \vcsp{} and \vcsp[\Gamma']{} are equivalent under polynomial-time reductions, where $\Gamma'$ is obtained by restricting each cost function in $\Gamma$ to $D \setminus \{d_1\}$.
    Moreover, the reductions do not change the primal graph of the instances.
\end{lemma}
\begin{proof}
    Restricting the unary functions that take value $1$ on one element and $0$ elsewhere shows that $\Gamma'$ is again a finite-valued soft-list language.
    For each $f\in\Gamma$ and each position $i$, fix a constant $c^{d_1,d_2}_{f,i}$ witnessing $d_1\additiveequiv d_2$ for $f$ and position $i$ (as defined in \cref{def:additive_equivalence}).
    Thus, changing the entry in position $i$ from $d_2$ to $d_1$ adds $c^{d_1,d_2}_{f,i}$ to the cost, independently of the other entries.

    Now, consider an instance $I=(V,\CC,t)$ of \vcsp{}.
    For each variable $v \in V$, we can determine whether it is better to assign $d_1$ or $d_2$ to $v$, independently of the assignments to the other variables.
    To do this, let $\CC_v$ be the constraints with $v$ in their scope, and calculate
    \[
        \Delta_v = \sum_{C \in \CC_v} \weight{C}
        \sum_{i:\,\scope{C}_i=v} c^{d_1,d_2}_{\costfun{C},i},
    \]
    where $\scope{C}_i = v$ expresses that $v$ is the $i$th element of the scope of $C$.
    The inner sum runs over all positions where $v$ occurs in the scope of $C$, which ensures that repeated occurrences of $v$ within a single constraint are counted separately.

    Consider a variable assignment $\pi$ with $\pi(v)=d_1$, and obtain $\pi'$ by changing only the assignment of $v$ to $d_2$.
    By \cref{def:additive_equivalence}, the cost of $\pi$ equals the cost of $\pi'$ plus $\Delta_v$.
    This follows by replacing the occurrences of $v$ one at a time and adding the corresponding weighted witnessing constants.
    Hence, we prefer $d_2$ if $\Delta_v \geq 0$, and $d_1$ if $\Delta_v<0$.

    In the output instance, we restrict all original cost functions to $D \setminus \{d_1\}$ and keep their scopes and weights.
    For each variable $v$ with $\Delta_v<0$, we add the soft list on $\{d_2\}$
    at~$v$ with weight $-\Delta_v$.
    Thus, the added cost at~$v$ is $-\Delta_v$
    unless $v$ is assigned~$d_2$, in which case it is zero.
    We set the target to
    \[
        t' = t + \sum_{v:\,\Delta_v<0} (-\Delta_v).
    \]
    For variables with $\Delta_v \geq 0$, we add no constraints.
    All added penalties are implemented by weighted unary constraints from $\Gamma'$ as described after the statement of \cref{def:soft_lists} in \cref{sec:vcsp_soft_lists}.

    To see that the reduction is correct, consider an assignment $\pi$ of the input instance of cost at most $t$.
    We may assume that whenever a variable is assigned one of $d_1,d_2$, it is assigned the preferred one.
    We then replace every occurrence of $d_1$ by $d_2$ to create $\pi'$.
    For a variable $v$ with $\Delta_v<0$, this increases the cost by $-\Delta_v$ if $v$ was assigned $d_1$.
    If $v$ was assigned any other element, it also increases the cost by $-\Delta_v$ due to the added unary penalty of $-\Delta_v$.
    Variables with $\Delta_v \geq 0$ contribute no change.
    Altogether, the cost of $\pi'$ is exactly the cost of $\pi$ plus $t'-t$.
    Therefore, $\pi'$ is an assignment of the output instance of cost at most $t'$.

    Conversely, take an output assignment $\pi'$ of cost at most $t'$.
    For each variable $v$ with $\Delta_v<0$ that is assigned $d_2$, change its assignment to $d_1$ to create assignment $\pi$.
    This decreases the cost of the original constraints by $-\Delta_v$.
    If such a variable is assigned another element, keep its assignment.
    Since the input instance does not contain its added unary penalty, again this decreases the cost by $-\Delta_v$.
    The resulting input assignment $\pi$ therefore has cost at most $t'-(t'-t)=t$.

    For the reduction in the other direction, that is, from \vcsp[\Gamma']{} to \vcsp{}, we only need to make sure that no variable is assigned the domain value $d_1$.
    Lift each constraint to its original cost function in $\Gamma$, and use a unary constraint to penalize assigning $d_1$ at any variable by $\max\{0,t\}+1$ (to keep the penalty positive when $t$ is negative), while not penalizing other assignments.
    An assignment to the input instance then gives an assignment to the output instance of the same cost.
    Since all costs are nonnegative, an assignment to the output instance of cost at most $t$ cannot use $d_1$.
    Hence, it again gives an assignment to the input instance of the same cost.

    It is clear from the description of the reductions that they run in polynomial time, and that they do not change the primal graph.
\end{proof}

We then directly obtain the following theorem.

\begin{theorem}
    \label{thm:soft_lists_alg}
    Let $D$ be a fixed finite domain and $\Gamma$ a finite-valued soft-list language over $D$.
    Let $\ccadditiveequiv$ be the number of equivalence classes of $\additiveequiv$.
    Given as input $I$ an instance of \vcsp{} and a tree decomposition $\mathcal T$ of its primal graph of width $\tw$, \vcsp{} can be solved in time $(\ccadditiveequiv)^{\tw}\cdot (|I|+|\mathcal T|)^{\Oh(1)}$.
\end{theorem}
\begin{proof}
    Use the reduction of \cref{thm:add_equiv_reduction} repeatedly until the domain of the obtained instance is $D'$, where $D'$ contains exactly one element of each equivalence class of $\additiveequiv$.
    The reductions preserve the primal graph, so the supplied tree decomposition remains valid.
    Every variable of the reduced instance has available set $D'$, of size $\ccadditiveequiv$.
    Thus \cref{thm:available_sets_dp} gives the claimed running time.
\end{proof}

\subsection{The Lower Bound}

Now, we show a lower bound matching \cref{thm:soft_lists_alg}.
Recall that we want to state our lower bounds in terms of a constant $c_{\Gamma}$ depending on the size of a domain subset on which we can express binary disequality.
For this purpose, let us formally define this constant.

\begin{definition}[Running time base $c_\Gamma$]
    \label{def:running_time_base}
    Let $D$ be a finite domain and $\Gamma$ be a constraint language over $D$.
    For a subset $D' \subseteq D$ a binary function $f$ with $f(x,y) = 0$ for $x,y \in D'$ with $x \neq y$, and $f(x,x) = 1$ for $x \in D'$ is a \emph{binary disequality function on $D'$}.
    We define $c_\Gamma$ to be the maximum cardinality of a subset $D' \subseteq D$ such that $\Gamma$ expresses a binary disequality function on $D'$.
\end{definition}

It is quite easy to obtain a lower bound using the running time base $c_\Gamma$, indeed if $c_\Gamma = 2$ a reduction from \textsc{Max Cut}, and if $c_\Gamma \geq 3$ a reduction from \textsc{$c_\Gamma$-Coloring} suffices.

\begin{lemma}
    \label{thm:lb_from_c_gamma}
    Let $D$ be a fixed finite domain and $\Gamma$ be a fixed constraint language over $D$.
    Let $D' \subseteq D$ be a set of size $c_\Gamma$ such that $\Gamma$ expresses a binary disequality function $f$ on $D'$.
    Assume that $\Gamma$ additionally expresses a restriction to $D'$, and that $c_\Gamma \geq 2$.
    Then, for every $\eps>0$, there is no algorithm solving \vcsp{} in time
    \[
        (c_\Gamma-\eps)^{\pw}\cdot (|I|+|\mathcal T|)^{\Oh(1)},
    \]
    where the input consists of an instance $I$ and a path decomposition $\mathcal T$ of its primal graph of width $\pw$, unless \pwseth{} is false.
    The same lower bound holds with pathwidth replaced by treewidth \mbox{under \twseth{}}.
\end{lemma}
\begin{proof}
    By shifting and scaling the expressed restriction to $D'$, fix $h\in\power{\Gamma}$ with $h(d)=0$ for $d\in D'$ and $h(d)\geq1$ for $d\notin D'$.

    Given a graph $G=(V,E)$ with $m=|E|$, construct an instance over $\Gamma\cup\{f,h\}$ with variable set $V$.
    For every edge $uv\in E$, add the constraint $(f,(u,v),1)$, choosing either ordering of its endpoints, and for every $v\in V$, add $(h,(v),m+1)$.
    An assignment where a variable uses a value outside $D'$ has cost at least $m+1$ due to this restriction, whereas every assignment over $D'$ has cost at most $m$.
    For an assignment $\pi:V\to D'$, the cost is exactly
    \[
        |\{uv\in E:\pi(u)=\pi(v)\}|.
    \]
    If $c_\Gamma=2$, a \textsc{Max Cut} instance asking for a cut of size at least $k$, where $0\leq k\leq m$, is therefore equivalent to the constructed instance with target $m-k$.
    If $c_\Gamma\geq3$, the graph $G$ has a proper $c_\Gamma$-coloring if and only if the constructed instance with target zero is a yes-instance.

    The constructed instance has the same primal graph as $G$.
    Since $\{f,h\}$ is a fixed finite subset of $\power{\Gamma}$, \cref{thm:expressive_power_reduction} replaces these functions by gadgets over $\Gamma$ in polynomial time, adjusts the target, and increases the width of the supplied path or tree decomposition by at most a constant.

    For $c_\Gamma=2$, the corresponding lower bound for \textsc{Max Cut} under \twseth{}
    follows from~\cite[Theorem~1]{IwataY15TWSETH}
    and the reduction in~\cite[Claim~5.2]{lampisTreewidthSETH}.
    For \pwseth{}, the lower bound
    follows from the \textsc{Max-2-SAT} lower bound in~\cite[Theorem~3.1]{lampisPrimalPathwidthSETH}, together with the reduction in~\cite[Claim~5.2]{lampisTreewidthSETH}, which also only increases the pathwidth by a constant.
    For $c_\Gamma\geq3$, the required lower bounds for \kcoloring[{c_\Gamma}]{} follow from~\cite[Lemma~5.1]{lampisPrimalPathwidthSETH} under \pwseth{} and from~\cite[Claims~5.3--5.4]{lampisTreewidthSETH} under \twseth{}.
    Thus, in both cases, an algorithm with the stated running time would contradict the corresponding hypothesis.
\end{proof}

Hence, the remainder of this section is dedicated to showing that $c_{\Gamma} = \ccadditiveequiv$, assuming that $c_{\Gamma} \geq 2$.
For this purpose, we show that all finite-valued cost functions on the domain remaining after preprocessing can be expressed, up to a common additive constant.
We begin by showing this for unary cost functions.

\begin{lemma}
    \label{thm:soft_lists_unary_costs}
    Let $D$ be a finite domain and $\Gamma$ a constraint language that expresses all soft lists.
    Then, $\Gamma$ expresses all finite-valued unary cost functions on all subsets of $D$.
\end{lemma}
\begin{proof}
    Let $B\subseteq D$ and $u:B\to\ratNnonneg$ be an arbitrary function.
    If $|D|\leq 1$, the function $h(d) = 0$ suffices since it is the same as $u$ up to a constant.

    So, assume $|D| \geq 2$.
    For each $d\in B$, let $h_d\in\power{\Gamma}$ be the soft list on $D\setminus\{d\}$, so $h_d(d)=1$ and $h_d(x)=0$ for $x\neq d$.
    Define
    \[
        h(x)=\sum_{d\in B}u(d)h_d(x),
    \]
    omitting terms with $u(d)=0$ in the sum.
    Then $h$ is a finite-valued nonnegative cost function expressing $u$ on $B$.
\end{proof}

We now use the work of Thapper and Živný~\cite{thapperComplexityFiniteValued2016} to establish that the problem we consider is polynomial-time solvable, or that $\Gamma$ expresses a finite-valued reversing move.
The following is a consequence of results in~\cite{thapperComplexityFiniteValued2016}, using the assumption that $\Gamma$ expresses all soft lists.

\begin{theorem}[{Follows from~\cite[Lemma~2.10,~Theorem~A.1 and Lemma~2.4]{thapperComplexityFiniteValued2016}}] \label{thm:soft_lists_reversing_move}
    Let $D$ be a fixed finite domain and $\Gamma$ be a finite-valued constraint language over $D$ expressing all soft lists.
    The problem \vcsp[\Gamma] can be solved in polynomial time, or there exist two distinct domain elements $d_1,d_2 \in D$ such that $\Gamma$ expresses a finite-valued reversing move on $\{d_1,d_2\}$.
\end{theorem}

Note that \cref{thm:soft_lists_reversing_move} above is indeed unconditionally true.
However, if $\pclass = \np$, then \vcsp[\Gamma]{} is always polynomial-time solvable and the first disjunct is always true.
Later, we will show that if $\Gamma$ expresses a finite-valued reversing move on a pair of domain elements, then we can express virtually all cost functions.

\begin{restatable}{lemma}{softListsExpressCosts}
    \label{thm:soft_lists_express_costs}
    Let $D$ be a finite domain and $\Gamma$ a finite-valued constraint language over $D$ expressing all soft lists.
    Let $D'\subseteq D$ contain exactly one element of each equivalence class of $\additiveequiv$.
    Assume that $\Gamma$ expresses a finite-valued reversing move on $\{a,b\}$ for some distinct $a,b\in D$.
    Then $|D'|\geq 2$ and $\Gamma$ expresses all finite-valued cost functions on $D'$.
\end{restatable}

We defer the proof of \cref{thm:soft_lists_express_costs} until the end of this subsection.
We are now ready to show that, assuming $c_{\Gamma} \geq 2$, we have $c_{\Gamma} = \ccadditiveequiv$.
A reversing move can only join different additive-equivalence classes, while \cref{thm:soft_lists_express_costs} supplies a single binary cost function with the required values on a set of representatives.

\thmSoftListBaseEquality*{}
\begin{proof}
    A function witnessing a set in \cref{def:running_time_base} is a reversing move on every two-element subset of that set.
    Since $\Gamma$ expresses all soft lists, neither of two distinct elements dominates the other.
    Hence a reversing move on $\{a,b\}$ witnesses interaction between $a$ and $b$.
    It follows directly from the definitions (and also from \cref{thm:additive_noninteraction} given later) that these elements cannot be additively equivalent.
    Thus any subset counted by $c_\Gamma$ contains at most one element from each additive-equivalence class, giving $c_\Gamma\leq \ccadditiveequiv$.

    Let $D'\subseteq D$ contain exactly one element of each additive-equivalence class.
    Since $c_\Gamma\geq2$ and $\Gamma$ is finite-valued, $\Gamma$ expresses a finite-valued reversing move on some pair of distinct elements.
    By \cref{thm:soft_lists_express_costs}, $\Gamma$ therefore expresses all finite-valued cost functions on $D'$, including one of the required form (where we use soft lists to ensure the values outside $D'$ are penalized sufficiently), and $c_\Gamma \geq \ccadditiveequiv$.
\end{proof}

Note that by \cref{thm:soft_lists_reversing_move} the problem is polynomial-time solvable when $c_\Gamma = 1$.
Hence, $c_\Gamma$ and $\ccadditiveequiv$ coincide for all cases that are not polynomial-time solvable, and our lower bound follows from \cref{thm:lb_from_c_gamma}.

\subparagraph*{Expressing cost functions.}
It remains to prove \cref{thm:soft_lists_express_costs}, that is, that we can express all finite-valued cost functions on the remaining domain.
First, we can observe that if all soft lists are available no element dominates another.

\begin{observation}
    \label{obs:soft_lists_no_domination}
    Let $D$ be a finite domain and $\Gamma$ a constraint language over $D$ expressing all soft lists.
    No element of $D$ dominates a distinct element.
\end{observation}

Domination consequently causes no additional restriction in the soft-list setting.
Now, we tie the notion of interaction to additive equivalence.

\begin{lemma}
    \label{thm:additive_noninteraction}
    Let $D$ be a finite domain and $\Gamma$ be a finite-valued constraint language over $D$ expressing a restriction to $\{d\}$ for each $d \in D$.
    Then, two distinct elements $d_1,d_2 \in D$ are additively equivalent if and only if they do not interact.
\end{lemma}
\begin{proof}
    First, assume that $d_1$ and $d_2$ are not additively equivalent.
    Then, there is a cost function $f \in \Gamma$ of arity $a$, a position $i \in \{1,\dots,a\}$, and domain elements $w_1,\dots,w_{a-1},z_1,\dots,z_{a-1}$ such that
    \begin{align*}
         & f(w_1,\dots,w_{i-1},d_1,w_i,\dots,w_{a-1})
        - f(w_1,\dots,w_{i-1},d_2,w_i,\dots,w_{a-1})             \\
         & \qquad\neq f(z_1,\dots,z_{i-1},d_1,z_i,\dots,z_{a-1})
        - f(z_1,\dots,z_{i-1},d_2,z_i,\dots,z_{a-1}).
    \end{align*}
    Consider the sequence of replacements, where the entries of the first tuple $w_1,\dots,w_{a-1}$ are changed to the entries of the second tuple $z_1,\dots,z_{a-1}$ one position at a time.
    Since the differences at the first and last tuples are different, at least one step changes this difference.
    The tuples before and after this step differ in only a single element.
    Hence, we can assume that $w_{j'} = z_{j'}$, apart from a single index $j$ in which they differ.

    Let $g(x,y)$ be obtained from $f$ by putting $x$ in position $i$, putting $y$ in the position corresponding to $w_j$ and $z_j$, and fixing every remaining argument to its common value in the two tuples.
    Then,
    \[
        g(d_1,w_j)-g(d_2,w_j) \neq g(d_1,z_j)-g(d_2,z_j).
    \]
    To express $g$, add for each fixed position an auxiliary variable with a restriction to its common value in both tuples, weighted so that its penalty for any other value
    exceeds $\max f - \min f$.
    Minimizing over the auxiliary variables then
    gives $g$ plus a constant, so $g\in\power{\Gamma}$.

    Due to the restrictions that $\Gamma$ expresses, neither of $d_1,d_2$ dominates the other.
    Thus, $g$ witnesses the interaction of $d_1$ and $d_2$.

    For the other direction, assume that $d_1$ and $d_2$ are additively equivalent, and consider any finite-valued binary cost function $g \in \power{\Gamma}$.
    Take a gadget expressing $g$, with distinguished variables $x$ and $y$.
    For any fixed assignment to $y$ and the auxiliary variables, changing $x$ from $d_2$ to $d_1$ changes the total cost by the same constant $\Delta_x$.
    This constant is obtained by summing the weighted witnessing constants $c^{d_1,d_2}_{f,i}$ over all occurrences of $x$ in the gadget's constraints, as in the proof of \cref{thm:add_equiv_reduction}.
    In particular, each constant used in the sum does not depend on the assignment to $y$ or the auxiliary variables.
    Minimizing over the auxiliary variables therefore preserves this difference, and any additive constant in the expression for $g$ cancels.
    Hence, for every $d' \in D$, we have
    \[
        g(d_1,d')-g(d_2,d')=\Delta_x.
    \]
    No such $g$ can witness an interaction, so $d_1$ and $d_2$ do not interact.
\end{proof}

We now show that essentially all moves exist.

\interactingPairsMove*
\begin{proof}
    Let $a',b'$ witness the interaction of $a,b$, and let $c',d'$ witness the interaction of $c,d$.
    Let $f,g\in\power{\Gamma}$ be finite-valued binary cost functions witnessing these two interactions, respectively.
    Thus, \cref{thm:interaction_move} gives $\{a,b\}\moves\{a',b'\}$ and $\{c,d\}\moves\{c',d'\}$.
    For each $x \in D$, define
    \[
        \Delta_1(x)=f(x,a')-f(x,b'),
        \qquad
        \Delta_2(x)=g(x,c')-g(x,d').
    \]
    By the choice of the witnesses, we have $\Delta_1(a) \neq \Delta_1(b)$ and $\Delta_2(c) \neq \Delta_2(d)$.

    If $\Delta_2(a) \neq \Delta_2(b)$, then $c',d'$ also witness the interaction of $a,b$.
    Hence, $\{a,b\} \moves \{c',d'\} \moves \{c,d\}$, which gives the desired move by \cref{thm:move_composition}.

    Otherwise, $\Delta_2(a)=\Delta_2(b)$.
    Since $\Delta_2(c) \neq \Delta_2(d)$, we can choose $e \in \{c,d\}$ such that $\Delta_2(e) \neq \Delta_2(a)$.
    Since $\Delta_1(a) \neq \Delta_1(b)$, we can also choose $r \in \{a,b\}$ such that $\Delta_1(r) \neq \Delta_1(e)$.
    We then have
    \[
        \Delta_1(r) \neq \Delta_1(e),
        \qquad
        \Delta_2(r)=\Delta_2(a) \neq \Delta_2(e).
    \]
    These inequalities imply $r\neq e$, and neither $r$ nor $e$ dominates the other by assumption.
    Thus, $a',b'$ and $c',d'$ both witness the interaction of $r,e$.
    Applying \cref{thm:interaction_move} and transposing moves where needed gives
    \[
        \{a,b\} \moves \{a',b'\} \moves \{r,e\}
        \moves \{c',d'\} \moves \{c,d\}.
    \]
    The moves compose and we have $\{a,b\} \moves \{c,d\}$.
    The moves obtained from \cref{thm:interaction_move} are finite-valued, and their compositions are also finite-valued by \cref{thm:move_composition}.
\end{proof}

\clearpage
We can now finish the deferred proof.
\softListsExpressCosts*
\begin{proof}
    Let $g\in\power{\Gamma}$ be a finite-valued reversing move on $P=\{a,b\}$.
    It satisfies $g(a,a)+g(b,b)>g(a,b)+g(b,a)$, and neither element dominates the other by \cref{obs:soft_lists_no_domination}.
    Thus $P$ is interacting.
    By \cref{thm:additive_noninteraction}, its elements are not additively equivalent, so $|D'|\geq 2$.
    Any two distinct elements of $D'$ belong to different equivalence classes, so $D'$ is pairwise interacting by the same lemma.

    By \cref{thm:soft_lists_unary_costs}, soft lists give all finite-valued unary cost functions on $P$ and on $D'$, as well as a restriction to $P$.
    The reversing move $g$ and the restriction to $P$ let us apply \cref{thm:boolean_costs} to express all finite-valued cost functions on $P$.
    \cref{thm:interacting_pairs_move} gives a finite-valued move from every pair of distinct elements in $D'$ to $P$.
    We can therefore apply \cref{thm:transfer_costs} to express all finite-valued cost functions on $D'$.
\end{proof}

\section{VCSP over Finite-Valued Hard-List Languages}
\label{sec:vcsp_hard_lists}
We now consider finite-valued hard-list languages, which need not contain all soft lists.
For this, we recall the formal definitions of hard lists and finite-valued hard-list languages.

\defHardLists*{}

\defHardListLanguage*{}

Our algorithm and lower bound will again be tied to properties of the available domain elements.
Unlike in the soft-list setting, distinct available elements may dominate one another.
An \emph{expressive set} combines interaction with the ability to express soft lists.

\expressiveSet*

If $\Gamma$ expresses all soft lists, condition~(2) is automatic, as in
\cref{sec:vcsp_soft_lists}.
Moreover, condition~(2) holds if and only if
$\Gamma$ expresses all finite-valued unary cost functions on~$D'$.
Indeed, for
every $S\subseteq D'$, the restriction of the soft list $h_S$ to~$D'$
is such a function.
Conversely, if $|D'|\ge2$, then for each $d\in D'$,
condition~(2) applied to the soft list on $D'\setminus\{d\}$ gives a
finite-valued unary $g_d\in\power{\Gamma}$ and $c_d\in\ratN$ with
$g_d=h_{D'\setminus\{d\}}+c_d$ on~$D'$.
For every $u\colon
    D'\to\ratNnonneg$, the function $\sum_{d\in D'}u(d)\,g_d$ then equals
$u+\sum_{d\in D'}u(d)\,c_d$ on~$D'$, as in the proof of
\cref{thm:soft_lists_unary_costs}.
If $|D'|=1$, every unary function
on~$D'$ is constant there, and condition~(2) applied to the soft list on~$D'$
gives a finite-valued unary function in $\power{\Gamma}$ that is constant
on~$D'$.

\subsection{The Algorithm}
First, we show that the running time of solving \vcsp{} on graphs of bounded treewidth can be tied to the size of expressive sets.
We preprocess our instance using the steps described next.

\subsubsection*{Preprocessing and Block Decomposition}

We begin with simple reductions that shrink the available sets of the variables, and then split the instance into independent subinstances according to the blocks of the move graph.

\paragraph{Removing dominated elements.}

First, we give a simple preprocessing algorithm that deletes dominated elements from the available sets of variables.
The upcoming deletion rules in \cref{thm:remove_dominated_elements,thm:remove_noninteracting_pairs} are special cases of soft neighborhood substitution~\cite{Cooper03Reduction,BistarelliFaltingsNeagu02,LecoutreRousselDehani12SNS,givryDeadEndElimination2013}: for a variable $v$ of an instance, a domain element $a$ can be discarded when there is another available element $b$ that is at least as good, regardless of the assignment to the neighboring variables.

\begin{lemma}
    \label{thm:remove_dominated_elements}
    Let $D$ be a finite domain and $\Gamma$ a finite, finite-valued hard-list language over $D$.
    Given an instance $I$ of \vcsp{}, one can in polynomial time add hard-list constraints to obtain an instance $I'$ of \vcsp{} with the same optimum cost and the same primal graph such that, for each variable, no available element is dominated by a distinct available element.
\end{lemma}
\begin{proof}
    Let $a,b$ be distinct available elements at a variable $v$, with $a$ dominating $b$.
    We show that changing the value of~$v$ from $b$ to~$a$ never increases
    the cost.
    The hard lists at~$v$ allow both $a$ and~$b$.
    Consider any other
    constraint at~$v$; its cost function~$f$ is finite-valued.
    Let $i$ be a
    position of~$f$.
    Suppose that, for some values at the other positions,
    $f$ is more expensive with~$a$ at position~$i$ than with~$b$.
    Fixing the
    other positions to these values by singleton hard lists expresses a
    finite-valued unary function in $\power{\Gamma}$ that is cheaper at~$b$
    than at~$a$, contradicting that $a$ dominates~$b$.
    Hence, for all values
    at the other positions, replacing $b$ by~$a$ at position~$i$ does not
    increase the cost of~$f$.
    Applying this to the occurrences of~$v$ one at
    a time proves the claim.

    Hence, adding a hard list at~$v$ that forbids~$b$ does not change the optimum cost.
    Repeating this while some variable has an available element dominated by
    a distinct available one removes one element each time, so it stops after
    at most $(|D|-1)\cdot|V|$ steps.
\end{proof}

\paragraph{Reduction to binary functions.}
We now introduce a useful lemma whose point is reducing properties of higher-arity functions to binary functions.
For a partition $S,T$ of $[m]$ and $\bar x,\bar y\in D^m$, we write
$\bar x_S\cdot\bar y_T$ for the tuple that agrees with $\bar x$ on~$S$
and with $\bar y$ on~$T$.
For $J\subseteq[m]$ and $\bar z\in D^m$, write $\bar z_J$ for the subtuple
of~$\bar z$ at the indices in~$J$, in increasing order.

\begin{lemma}\label{thm:binary-reduction}
    Let $D$ be a finite domain, $\Gamma$ a finite-valued hard-list language over~$D$, and $g\in\power{\Gamma}$ a finite-valued
    cost function of arity~$m$.
    Let $S,T$ be a partition of $[m]$ and
    $\bar x,\bar y\in D^m$ with
    \[
        g(\bar x)+g(\bar y)<g(\bar x_S\cdot\bar y_T)+g(\bar y_S\cdot\bar x_T).
    \]
    Then there are $i\in S$ and $j\in T$ with $x_i\ne y_i$ and $x_j\ne y_j$,
    and a finite-valued binary $h\in\power{\Gamma}$ such that
    \[
        h(x_i,x_j)+h(y_i,y_j)<h(x_i,y_j)+h(y_i,x_j).
    \]
    The same holds with both inequalities reversed.
\end{lemma}

\begin{proof}
    The first claim is~\cite[Lemma~12]{FZ17}.
    Since $g$ is finite-valued,
    the proof there only fixes coordinates to entries of $\bar x$
    or~$\bar y$.
    We express this using singleton hard lists, so the binary
    function it produces is a finite-valued $h\in\power{\Gamma}$.
    The strict
    inequality forces $x_i\ne y_i$ and $x_j\ne y_j$.
    For the reversed
    inequalities, apply the first claim to $\bar x_S\cdot\bar y_T$ and
    $\bar y_S\cdot\bar x_T$.
\end{proof}

\paragraph{Ensuring interaction.}

Next, we extend our preprocessing to make sure the available sets are pairwise interacting.
We reuse the cost comparison argument for additive equivalence, applied to the currently available assignments.

\begin{lemma}
    \label{thm:remove_noninteracting_pairs}
    Let $D$ be a finite domain and $\Gamma$ a finite, finite-valued hard-list language over $D$.
    Given an instance $I$ of \vcsp{} in which every variable has a nonempty set of available elements, one can in polynomial time add hard-list constraints to obtain an instance $I'$ with the same optimum cost and primal graph such that every variable has a nonempty, pairwise interacting set of available elements.
\end{lemma}
\begin{proof}
    First apply \cref{thm:remove_dominated_elements} to remove elements that are dominated by another available element from each available set.
    At a variable $v$, identify repeated occurrences of each variable in every incident non-hard-list constraint $C$, to obtain a finite-valued cost function $g_C(x,\bar y)\in\power{\Gamma}$ with $x$ representing $v$.
    Test whether distinct available elements $a,b$ make
    \[
        g_C(a,\bar y)-g_C(b,\bar y)
    \]
    constant over all available assignments $\bar y$, for each such constraint.
    If this is indeed the case, like in the cost comparison in the proof of \cref{thm:add_equiv_reduction}, summing these constants with the constraint weights determines whether assigning $a$ or $b$ is more expensive, and this is independent of the assignment to the other variables.
    Retain the cheaper choice and forbid the other by a hard list.
    This preserves the optimum, since both choices were available.
    Repeat this operation while such a deletion is possible.

    To show that the remaining available sets are pairwise interacting, suppose that distinct available elements $a,b$ at $v$ do not interact.
    If $g_C(a,\bar y)-g_C(b,\bar y)$ were nonconstant for some~$C$, that is, different for two available tuples $\bar y,\bar y'$, then \cref{thm:binary-reduction}, applied to $(a,\bar y)$ and $(b,\bar y')$ with $S$ consisting of the first position, would give a
    finite-valued binary function $q\in\power{\Gamma}$ and distinct available values $c,d$ at another variable with \[ q(a,c)-q(b,c)\ne q(a,d)-q(b,d). \]
    The initial preprocessing using \cref{thm:remove_dominated_elements} excludes domination on both available pairs, so $q$ would witness interaction of $a,b$.
    Hence all these differences are constant, and the preprocessing would have
    removed $a$ or $b$ from $L_v$.

    There are at most $|D|-1$ deletions per variable, and $\Gamma, D$ are fixed and finite.
    Hence the procedure runs in polynomial time.
    As only unary constraints are added the primal graph is unchanged.
\end{proof}

\paragraph{The move graph.}
By \cref{thm:interaction_move}, every interaction witness yields a finite-valued move to one ordering of its witness pair.
We capture these moves in a graph on ordered pairs that we call the \emph{move graph}.

\begin{definition}[Move graph]
    \label{def:move_graph}
    Let $D$ be a finite domain and $\Gamma$ a constraint language over $D$.
    The \emph{move graph} $G_\Gamma$ (of $\Gamma$) has a vertex $(a,b)$ for each ordering of an interacting pair $\{a,b\}$.
    There is an edge between $(a,b)$ and $(c,d)$ if a finite-valued move from $(a,b)$ to $(c,d)$ belongs to $\power{\Gamma}$.
    Loops are allowed.
\end{definition}

Swapping the two arguments of a finite-valued move gives, by \cref{thm:move_composition}, a finite-valued move in the reverse direction, so the move graph is undirected.
The second part of \cref{thm:move_composition} lets us compose moves along walks.

\begin{observation}\label{obs:move_composition}
    Let $D$ be a finite domain and $\Gamma$ a finite, finite-valued hard-list language over~$D$.
    If $(a,b)$ and $(c,d)$ are in the same connected component of the move graph~$G_\Gamma$, then there is a finite-valued move from $(a,b)$ to $(c,d)$ in $\power{\Gamma}$.
\end{observation}

\begin{proof}
    Compose the moves along a walk from $(a,b)$ to $(c,d)$ in~$G_\Gamma$, using
    \cref{thm:move_composition} with the hard list on each intermediate pair as
    the restriction.
    If $(a,b)=(c,d)$, use a walk to a neighbor and back.
    A
    neighbor exists because $\{a,b\}$ is interacting, so
    \cref{thm:interaction_move} gives a finite-valued move from $(a,b)$ to
    an ordering of its witness pair, which is itself interacting.
\end{proof}

\paragraph{Blocks of the move graph.}
When $(a,b)\moves (c,d)$, we also have $(b,a)\moves (d,c)$, witnessed by the same move function (see the discussion in \cref{sec:interaction_moves}).
Naturally, either $(a,b)$ is in the same connected component of the move graph as $(b,a)$, or in a different one.
It turns out that this is a very crucial difference.
Indeed, a finite-valued reversing move on $\{a,b\}$ can be expressed only if $(a,b)$ and $(b,a)$ are in the same connected component.

Consider some connected component $C$ of the move graph, and let $(a,b)$ be a vertex of the connected component.
If $(b,a)$ is also part of $C$, then for all vertices $(a',b')$ of $C$ we have
that also $(b',a')$ is part of $C$ by the preceding note.
Otherwise, let $C'$ be the connected component of the move graph that contains $(b,a)$.
Then, again by the preceding note, we have for all vertices $(a',b')$ of $C$ that $(b',a')$ is part of $C'$.
We define \emph{blocks} of the move graph based on this observation.

For a vertex $(a,b)$ of the move graph let $\rho(a,b) = (b,a)$ be the vertex obtained by reversing the pair.
Similarly, for a set of vertices $S$ of the move graph let $\rho(S)$ be the set obtained from $S$ by applying $\rho$ to each element of $S$.

\begin{definition}[Blocks of the move graph]
    \label{def:blocks}
    Let $D$ be a finite domain and $\Gamma$ a constraint language over $D$.
    Let $G_\Gamma$ be the move graph of $\Gamma$.
    A \emph{block} of $G_\Gamma$ is a set $B=C\cup\rho(C)$, where $C$ is the vertex set of a connected component of $G_\Gamma$.
    It is \emph{reversing} if $C=\rho(C)$ and \emph{nonreversing} otherwise.
\end{definition}

In a reversing block, \cref{obs:move_composition} gives a finite-valued reversing move on every pair in the block.
In a nonreversing block, each component contains exactly one ordering of every unordered pair.
Blocks give a natural partition of the vertex set of the move graph, and we can assign unordered pairs to unique blocks.

\begin{definition}[Blocks of pairs]\label{def:blocks_of_pairs}
    Let $D$ be a finite domain, $\Gamma$ a constraint language over $D$, and $G_\Gamma$ the move graph.
    For a vertex $(a,b) \in V(G_\Gamma)$ the \emph{block of $(a,b)$} is the block of the move graph that contains $(a,b)$.
    For an interacting pair $\{a,b\}$ the \emph{block of $\{a,b\}$} is the block of $(a,b)$.
    When $B$ is the block of $(a,b)$ $(\{a,b\})$ we also say that $(a,b)$ $(\{a,b\})$ \emph{belongs} to $B$.
\end{definition}

Next, we show that, as for pairs in \cref{def:blocks_of_pairs}, every pairwise interacting set of at least two elements can be assigned to a unique block.

\availableSetBlock*
\begin{proof}
    The claim holds for $|L|=2$ by the definition of a block.
    Let
    $a,b,c\in L$ be distinct, and let $g$ and $\{s,t\}$ witness the
    interaction of $a,b$ (transposing $g$ shows that $\{s,t\}$ is interacting).
    Writing $\Delta(x,y)=g(x,s)-g(y,s)-g(x,t)+g(y,t)$, we have
    $0\ne\Delta(a,b)=\Delta(a,c)+\Delta(c,b)$.
    Hence $g$ and $\{s,t\}$ also
    witness the interaction of $a,c$ or of $c,b$.
    By
    \cref{thm:interaction_move}, the pair $\{s,t\}$ lies in the block of
    $\{a,b\}$ and in the block of $\{a,c\}$ or of $\{c,b\}$.
    So each of the
    three pairs on $\{a,b,c\}$ shares its block with another of them, and all
    three belong to one block.
    Finally, two disjoint pairs $\{a,b\}$ and
    $\{c,d\}$ in~$L$ both share a block with $\{a,c\}$.
\end{proof}

This lemma justifies extending our notion of pairs belonging to a block to sets belonging to a block.

\begin{definition}[Blocks of sets]\label{def:blocks_of_sets}
    Let $D$ be a finite domain, $\Gamma$ a constraint language over $D$, $G_\Gamma$ its move graph, and $S$ a pairwise interacting set of elements of $D$.
    If $|S| \leq 1$, then $S$ does not belong to a block.
    If $|S| \geq 2$, let $a,b$ be two distinct elements of $S$; then $S$
    \emph{belongs} to the block of $(a,b)$.
\end{definition}

By \cref{thm:available_set_block} the block of a set is well-defined and the choice of the elements $a,b$ does not matter.
Since $\Gamma$ is fixed, the move graph, its blocks, and the block of every
pairwise interacting subset of~$D$ are fixed finite data that our algorithms
may use.

\paragraph{Decomposing instances.}

The purpose of blocks now is that they give us a clean decomposition of any instance of \vcsp{}.
Concretely, we can decompose a given instance into multiple instances in which every available set of size at least two belongs to a single block of the move graph.

The preceding preprocessing leaves every available set nonempty and pairwise interacting.
We now separate the instance according to the blocks of the move graph.\footnote{Similar decompositions into sums of functions of smaller arity appear, for example, in~\cite{BacchusGrove95,KuoSloanWasilkowskiWozniakowski10,HelaouiNaanaa13}.}

\begin{lemma}
    \label{thm:independent_block_instances}
    Let $D$ be a fixed finite domain and $\Gamma$ a fixed finite, finite-valued hard-list language over $D$.
    Let $I$ be an instance of \vcsp{} in which every variable has a nonempty, pairwise interacting set of available elements.
    For each block $B$ of the move graph $G_\Gamma$, let $V_B$ be the set of variables of $I$ whose available sets belong to $B$.

    One can in polynomial time construct a constant $c\in\ratN$ and, for each block $B$ with $V_B\neq\emptyset$, an instance $I_B$ of \vcsp[\Gamma]{} such that
    \[
        \Opt(I)=c+\sum_B\Opt(I_B),
    \]
    where the sum ranges over blocks with $V_B\neq\emptyset$.
    We call the instances $I_B$ the \emph{block instances} of this decomposition.
    Each variable of $I_B$ has an available set that either belongs to $B$ or has size one.
    A supplied tree or path decomposition of the primal graph of $I$ of width $k$ yields decompositions of the same type and width at most $k+\Oh(1)$ for all $I_B$ in polynomial time.
\end{lemma}
\begin{proof}
    For each variable $v$ of $I$, let $L_v$ be its available set.
    We will decompose each non-hard-list constraint $C$ into a sum of constraints whose scopes belong only to a single block.

    To do this, fix a non-hard-list constraint $C$ of the instance.
    Let $\bar v_C=(v_1,\dots,v_m)$ list the distinct variables of its scope in order of first occurrence, and let $g_C:D^m\to\ratNnonneg$ be its cost function after identifying repeated variables.
    Clearly the resulting function $g_C$ is in $\power{\Gamma}$.
    Write
    $\mathcal B_C=\{B\mid v_i\in V_B\text{ for some } i\in [m]\}$ for the set of
    blocks the variables in the scope of the constraint belong to, and $q_C=|\mathcal B_C|$.
    Choose an arbitrary ``reference'' tuple $\bar p_C=(p_1,\dots,p_m)\in L_{v_1}\times\cdots\times L_{v_m}$.
    Such a tuple exists because all available sets are nonempty by assumption.
    For each $B\in\mathcal B_C$, put $J_{C,B}=\{i\in[m] \mid v_i \in V_B\}$.
    That is, the sets $J_{C,B}$ partition the indices of variables with available sets of size at least two according to their blocks.

    For each block
    $B\in\mathcal B_C$, let $g_{C,B}\colon D^{|J_{C,B}|}\to\ratNnonneg$
    be obtained from $g_C$ by fixing every position outside $J_{C,B}$ to its
    value in~$\bar p_C$; that is,
    $g_{C,B}(\bar z_{J_{C,B}})
        =g_C\bigl(\bar z_{J_{C,B}}\cdot(\bar p_C)_{[m]\setminus J_{C,B}}\bigr)$
    for $\bar z\in D^m$.
    Set $\alpha_C=(1-q_C)\,g_C(\bar p_C)$.
    We show that,
    for every $\bar z\in L_{v_1}\times\dots\times L_{v_m}$,
    \begin{equation}\label{eq:separation_blocks}
        g_C(\bar z)=\alpha_C+\sum_{B\in\mathcal B_C}g_{C,B}(\bar z_{J_{C,B}}).
    \end{equation}

    Fix $B\in\mathcal B_C$, and let $S=J_{C,B}$ and $T=[m]\setminus S$.
    For
    all $\bar x,\bar y\in L_{v_1}\times\dots\times L_{v_m}$, we have
    \begin{equation}\label{eq:block-modular}
        g_C(\bar x)+g_C(\bar y)=g_C(\bar x_S\cdot\bar y_T)+g_C(\bar y_S\cdot\bar x_T).
    \end{equation}
    Otherwise, \cref{thm:binary-reduction} gives $i\in S$, $j\in T$, and
    a finite-valued binary $h\in\power{\Gamma}$ with
    $h(x_i,x_j)-h(y_i,x_j)\ne h(x_i,y_j)-h(y_i,y_j)$.
    By assumption, neither element of $x_i,y_i$ and neither element of $x_j,y_j$ dominates the other.
    Hence, $x_i,y_i$ interact, witnessed by $h$ and $x_j,y_j$ (\cref{def:interaction}).
    By \cref{thm:interaction_move},
    we then have that $\{x_i,y_i\}$ belongs to the same block as $\{x_j,y_j\}$, contradicting $v_i\in V_B$ and
    $v_j\notin V_B$.

    Taking $\bar x=\bar z$ and $\bar y=\bar p_C$ in~\eqref{eq:block-modular}
    gives $g_C(\bar z)=g_{C,B}(\bar z_S)+g_C(\bar z')-g_C(\bar p_C)$, where
    $\bar z'=(\bar p_C)_S\cdot\bar z_T$.
    Applying this to the blocks of
    $\mathcal B_C$ one at a time, and noting that $\bar z$ agrees with
    $\bar p_C$ outside the sets $J_{C,B}$, we obtain
    \[
        g_C(\bar z)=g_C(\bar p_C)+\sum_{B\in\mathcal B_C}
        \bigl(g_{C,B}(\bar z_{J_{C,B}})-g_C(\bar p_C)\bigr),
    \]
    which is Equation~\eqref{eq:separation_blocks}.
    This includes the case $q_C=0$.

    Perform this construction for every non-hard-list constraint $C$, and let $c=\sum_C\weight{C}\alpha_C$.
    For each $B$ with $V_B\neq\emptyset$, construct an intermediate weighted constraint system $J_B$ on the variables $V_B$, retaining the hard lists of $I$ on these variables.
    For every $C$ with $B\in\mathcal B_C$, add $(g_{C,B},\bar v_{C,B},\weight{C})$ to $J_B$, where $\bar v_{C,B}$ is the subtuple of $\bar v_C$ indexed by $J_{C,B}$.
    For every available assignment $\pi$ of $I$, write $\pi_B$ for its restriction to $V_B$.
    Then, \cref{eq:separation_blocks} gives
    \[
        \cost_I(\pi)
        =c+\sum_B\cost_{J_B}(\pi_B).
    \]
    Conversely, assignments to the disjoint sets $V_B$ combine with the unique available values of all other original variables.
    Taking optimal assignments gives $\Opt(I)=c+\sum_B\Opt(J_B)$.

    Every scope of $J_B$ is contained in an original constraint scope.
    Intersecting the bags of a supplied tree or path decomposition with $V_B$ therefore gives a decomposition of $J_B$ of the same type and no greater width.

    To obtain the desired instance, it remains to introduce gadgets for the functions $g_{C,B}$.
    This is easy to achieve while only increasing the width of the provided decomposition by a constant, because each function $g_{C,B}$ is obtained from a function of $\Gamma$ by identifying variables, and fixing variables to domain values using hard lists.
\end{proof}

\subsubsection*{Solving Instances of Nonreversing Blocks.}

We now explain how instances of nonreversing blocks can be solved in polynomial time.
As in~\cite[Section~5]{KZ13}, a nonreversing block yields a symmetric
tournament pair (STP) multimorphism, so the corresponding instances can
be solved by the algorithm of Cohen, Cooper, and Jeavons~\cite{CCJ08},
which relies on a reduction to submodular function minimization.

\begin{lemma}\label{thm:solve_nonreversing_block}
    Let $D$ be a fixed finite domain and $\Gamma$ a fixed finite, finite-valued hard-list language over~$D$.
    Let $I$ be an instance of
    \vcsp{} in which every variable has a nonempty, pairwise
    interacting set of available elements.
    If all available sets of size at
    least two belong to one nonreversing block of $G_\Gamma$, then an
    optimal assignment of~$I$ can be found in polynomial time.
\end{lemma}

\begin{proof}
    If all available sets are singletons, the claim is trivial.
    Let $B=C\cup\rho(C)$ be the block.
    Every finite-valued binary
    $h\in\power{\Gamma}$ satisfies
    \begin{equation}\label{eq:orientation}
        h(a,s)+h(b,t)\le h(a,t)+h(b,s)\qquad\text{for all }(a,b),(s,t)\in C.
    \end{equation}
    Otherwise, \cref{thm:interaction_move} gives a move from
    $(a,b)\in C$ to $(t,s)\in\rho(C)$, so $C=\rho(C)$, contradicting that
    $B$ is nonreversing; compare~\cite[Lemma~5.3]{KZ13}.
    Every pair
    belonging to~$B$ thus has exactly one ordering in~$C$.
    Hence there is an
    STP $\langle\sqcap,\sqcup\rangle$ on~$D$, that is, a pair of commutative
    operations with $\{a\sqcap b,a\sqcup b\}=\{a,b\}$, such that
    $a\sqcap b=a$ whenever $(a,b)\in C$.

    Identify repeated variables in each non-hard-list constraint, and replace
    its finite-valued cost function $g\in\power{\Gamma}$ by the function~$f$
    that agrees with~$g$ on the product $L_1\times\dots\times L_k$ of the
    available sets and is $\infty$ elsewhere.
    This preserves all available
    assignments and their costs, and only finitely many such~$f$ arise.
    Hard
    lists admit every conservative pair, so by~\cite{CCJ08} it suffices to
    show $f(\bar x\sqcap\bar y)+f(\bar x\sqcup\bar y)\le f(\bar x)+f(\bar y)$
    for $\bar x,\bar y\in L_1\times\dots\times L_k$, where both $\sqcap$ and $\sqcup$ act componentwise on tuples; i.e., $\bar x\sqcap \bar y=(x_1\sqcap y_1,\ldots,x_k\sqcap y_k)$ and
    $\bar x\sqcup \bar y=(x_1\sqcup y_1,\ldots,x_k\sqcup y_k)$, where $\bar x=(x_1,\ldots,x_k)$ and $\bar y=(y_1,\ldots, y_k)$.

    Let $T=\{j: x_j\ne y_j,\ x_j\sqcap y_j=y_j\}$ and $S=[k]\setminus T$, so
    that $\bar x\sqcap\bar y=\bar x_S\cdot\bar y_T$ and
    $\bar x\sqcup\bar y=\bar y_S\cdot\bar x_T$.
    If the inequality failed,
    \cref{thm:binary-reduction}, applied to~$g$, which agrees with~$f$
    on these four tuples, would give $i\in S$ and $j\in T$ with
    $x_i\ne y_i$ and $x_j\ne y_j$, and a finite-valued binary
    $h\in\power{\Gamma}$ with
    $h(x_i,y_j)+h(y_i,x_j)>h(x_i,x_j)+h(y_i,y_j)$.
    So $L_i$ and $L_j$ have
    size at least two, and the pairs $\{x_i,y_i\}$ and $\{x_j,y_j\}$ belong
    to~$B$ (\cref{def:blocks_of_sets}).
    The choice of $S$ and~$T$ now
    gives $(x_i,y_i),(y_j,x_j)\in C$, contradicting~\eqref{eq:orientation}.
\end{proof}

Together with \cref{thm:independent_block_instances}, this lets us compute
the contribution of every nonreversing block to the optimum in polynomial
time.

\subsubsection*{Expressing Cost Functions on Reversing Blocks}

Before proceeding with our approach for reversing blocks, we elaborate on how reversing blocks help us to express cost functions, and establish an alternative characterization of $c_\Gamma$ revolving around expressive sets.
First, we establish that we can express cost functions on certain set products, assuming that we can express finite-valued unary cost functions.

\begin{corollary}
    \label{cor:reversing_block_costs}
    Let $D$ be a finite domain and $\Gamma$ a finite, finite-valued hard-list language over $D$.
    Let $B$ be a reversing block of $G_\Gamma$, let $r\geq 1$, and let $S_1,\dots,S_r\subseteq D$ be nonempty subsets.
    Suppose that every $S_i$ of size at least two is pairwise interacting and belongs to $B$, and that $\Gamma$ expresses all finite-valued unary cost functions on each $S_i$.
    For every $f:S_1\times\cdots\times S_r\to\ratNnonneg$, there are a finite-valued $g\in\power{\Gamma}$ and a constant $c\in\ratN$ with $g=f+c$ on $S_1\times\cdots\times S_r$.
\end{corollary}
\begin{proof}
    Fix an interacting pair $P=\{p,q\}$ in $B$.
    Neither $p$ nor $q$ dominates the other, so \cref{thm:undominated_pair_unaries} shows that $\Gamma$ expresses all finite-valued unary cost functions on $P$.

    Since $B$ is reversing, \cref{obs:move_composition} gives a finite-valued reversing move on $P$.
    The hard list on $P$ is zero on $P$ and infinite outside it.
    We can therefore apply \cref{thm:boolean_costs} to express all finite-valued cost functions on $P$.

    Every pair of distinct elements in any set $S_i$ belongs to $B$, so a walk in this block gives a finite-valued move from that pair to $P$ by \cref{obs:move_composition}.
    Together with the assumed unary expressibility on each $S_i$, these moves and the hard list on $P$ satisfy the hypotheses of \cref{thm:transfer_costs}, which proves the claim.
\end{proof}

This allows us to characterize the value $c_\Gamma$ in terms of the size of expressive sets belonging to reversing blocks.

\thmHardListBaseEquality*{}
\begin{proof}
    Let $c' = \max\bigl(\{1\}\cup
        \{|S| \mid S\subseteq D\text{ is expressive and belongs to a reversing block of }G_\Gamma\}\bigr)$.
    Both $c_\Gamma$ and $c'$ are at least one: this holds for $c'$ by definition, and for $c_\Gamma$ because the constant binary function with value one is expressible.
    Let $S$ be an expressive set belonging to a reversing block.
    Expressiveness gives all finite-valued unary cost functions on $S$, so by \cref{cor:reversing_block_costs} (and using hard lists to penalize values outside $S$ if necessary) we have $c'\leq c_\Gamma$.

    For the reverse inequality, assume that $c_\Gamma\geq2$, since otherwise $c'\geq1$ suffices.
    By \cref{def:running_time_base}, there exists some set $S \subseteq D$ of size $c_\Gamma$ such that $\Gamma$ expresses a binary disequality function $f$ on $S$.
    In a gadget for $f$, remove all hard lists on the two distinguished variables (if such hard lists exist).
    This does not change the values on $S^2$.
    The resulting binary function is finite on all of $D^2$ since all auxiliary variables can take at least one value not forbidden by a hard list.
    By adding an additive constant if necessary we obtain a nonnegative finite-valued function $\widehat f\in\power{\Gamma}$ that agrees with $f$ on $S^2$ up to a constant $\kappa$.

    For each $a\in S$, pinning the second argument to $a$ using a hard list gives the finite-valued unary function $h_a(x)=\widehat f(x,a)$.
    On $S$, this function has value $\kappa+1$ at $a$ and $\kappa$ at every other element.
    Nonnegative weighted sums of these functions express all finite-valued unary costs on $S$ up to a constant.
    In particular, no two distinct elements of $S$ dominate one another, and $\widehat f$ witnesses interaction between every pair.
    Thus $S$ is expressive and, by \cref{thm:available_set_block}, belongs to one block.
    This block is reversing because $\widehat f$ is a reversing move on every two-element subset of $S$.
    Hence $|S|\leq c'$, proving the reverse inequality.
\end{proof}

\subsubsection*{Removing Redundant Elements of Reversing Blocks.}
Since we can solve nonreversing blocks in polynomial time, it remains to explain how our algorithm handles reversing blocks.
It will turn out that these are hard.
Nevertheless, we can do further nontrivial preprocessing, with the goal of ensuring that in the end, each available set of a variable belonging to a reversing block is expressive.
For this purpose, we define when an available element is locally necessary.
Intuitively, an element is locally necessary at a variable if some assignment to its neighbors makes it strictly cheaper than every other available element at that variable.

\begin{definition}[Locally necessary]
    \label{def:locally_necessary}
    Let $D$ be a finite domain and $\Gamma$ a finite-valued hard-list language over $D$.
    Consider an instance $I=(V,\CC,t)$ of \vcsp{}, and let $L_u$ be the nonempty set of available elements at each variable $u\in V$.
    Fix a variable $v\in V$, and let $N_v$ be its neighbors in the primal graph.
    Let $\CC_v$ be the family of finite-valued constraints involving $v$.
    For an assignment $\pi:N_v\to D$ with $\pi(u)\in L_u$ for every $u\in N_v$ and an element $d\in L_v$, let $\pi_d$ extend $\pi$ by assigning $d$ to $v$, and put
    \[
        h_v(d,\pi)=\sum_{C\in\CC_v}\weight{C}\costfun{C}\bigl(\pi_d(\scope{C})\bigr).
    \]
    An element $a\in L_v$ is \emph{locally necessary at $v$} if there is such an assignment $\pi$ for which, for every $b\in L_v\setminus\{a\}$,
    \[
        h_v(a,\pi)<h_v(b,\pi).
    \]
    In other words, $a$ is the unique minimizer over $L_v$ for some available assignment to the neighbors of $v$.
\end{definition}

An element that is not locally necessary can be forbidden at $v$ without changing the optimum: for every available assignment to the neighbors, another element is at least as good.
Note that the replacement element may depend on the neighbor assignment, as in related context-dependent substitution rules for CSPs~\cite{BordeauxCadoliMancini08,likitvivatanavongEliminatingRedundancyCSPs2013}.
For unary and binary cost functions, local necessity corresponds to a condition used by Koster, van Hoesel, and Kolen~\cite{KosterVanHoeselKolen02} for domain-reduction purposes in their treewidth-based algorithm.
For unary and binary cost functions, failure of local necessity also corresponds to bottom-line dead-end elimination~\cite{DesmetDeMaeyerLasters97,pierceConformationalSplitting2000}, and more generally represents an instance of improving mappings~\cite{shekhovtsovHigherOrderPersistency2016}.

We can compute whether an element is locally necessary by using the following mixed integer linear program (MILP).

\begin{mixedintegerprogram}[Local necessity test]
    \label{milp:local_necessity}
    Let $D$ be a finite domain and $\Gamma$ a finite-valued hard-list language over $D$.
    Let $I=(V,\CC,t)$ be an instance of \vcsp{}, and let $L_u$ be the nonempty set of available elements at each variable $u\in V$.
    Fix a variable $v\in V$ and an element $a\in L_v$, and let $N_v$ be the neighbors of $v$ in the primal graph.
    Let $\CC_v$ be the family of finite-valued constraints involving $v$.
    For each $C\in\CC_v$, identify repeated variables in its scope and place $v$ first, obtaining a finite-valued cost function $g_C(d,\bar y)$ on distinct variables.
    Let $u_{C,1},\dots,u_{C,k_C}$ be the other variables in the constraint scope, and put
    \[
        Y_C=L_{u_{C,1}}\times\cdots\times L_{u_{C,k_C}}.
    \]
    If $k_C=0$, then $Y_C$ consists of the empty tuple.

    We will use a binary indicator variable $q_{u,d}$ for assigning $d\in L_u$ to each neighbor $u\in N_v$, and a binary indicator variable $p_{C,\bar y}$ for choosing $\bar y\in Y_C$ in each constraint $C\in\CC_v$.
    For each element $b\in L_v\setminus\{a\}$, define the cost difference
    \[
        d_a(b,p)=\sum_{C\in\CC_v}\weight{C}
        \sum_{\bar y\in Y_C}p_{C,\bar y}
        \bigl(g_C(b,\bar y)-g_C(a,\bar y)\bigr).
    \]
    This expression measures the cost difference between assigning $b$ to the variable $v$ and assigning $a$ to the variable.
    It is linear in the tuple indicators $p_{C,\bar y}$, with rational coefficients.
    We also use a rational variable $\eps$ for the common cost gap.
    The mixed integer linear program is then
    \crefalias{equation}{constraint}
    \begin{align}
         & \text{maximize}\quad\eps\nonumber                                                                                                                           \\
        \text{subject to}\quad
         & \sum_{d\in L_u}q_{u,d}=1
         &                                                    & \text{for every }u\in N_v,\label{eq:local_necessity_neighbors}                                         \\
         & \sum_{\bar y\in Y_C}p_{C,\bar y}=1
         &                                                    & \text{for every }C\in\CC_v,\label{eq:local_necessity_tuples}                                           \\
         & \sum_{\substack{\bar y\in Y_C,                                                                                                                              \\y_j=d}}p_{C,\bar y}=q_{u_{C,j},d}
         &                                                    & \text{for every }C\in\CC_v,\ 1\leq j\leq k_C,\ d\in L_{u_{C,j}},\label{eq:local_necessity_consistency} \\
         & \eps \leq d_a(b,p)
         &                                                    & \text{for every }b\in L_v\setminus\{a\},\label{eq:local_necessity_margin}                              \\
         & \eps \leq 1,\label{eq:local_necessity_upper_bound}                                                                                                          \\
         & q_{u,d}\in\{0,1\}
         &                                                    & \text{for every }u\in N_v,\ d\in L_u,\label{eq:local_necessity_neighbor_integrality}                   \\
         & p_{C,\bar y}\in\{0,1\}
         &                                                    & \text{for every }C\in\CC_v,\ \bar y\in Y_C.\label{eq:local_necessity_tuple_integrality}
    \end{align}
\end{mixedintegerprogram}

The optimum of this mixed integer linear program is positive if and only if element $a$ is locally necessary.
To see this, let us elaborate on the chosen constraints.
Together with \cref{eq:local_necessity_neighbor_integrality}, \cref{eq:local_necessity_neighbors} ensures that each neighbor of $v$ in the primal graph is assigned exactly one domain element.
Similarly, \cref{eq:local_necessity_tuples,eq:local_necessity_tuple_integrality} ensure that each constraint scope has a unique assignment.
\cref{eq:local_necessity_consistency} ensures that the assignments to the constraint scopes match the assignments to the neighbors of $v$.
\cref{eq:local_necessity_margin} guarantees that $\eps$ is positive only if, for the chosen assignment to the neighbors of $v$, assigning $a$ is strictly better than assigning any other element in the available set of $v$.
Finally, \cref{eq:local_necessity_upper_bound} is added to keep the objective bounded in all cases, without affecting whether there exists a positive optimum (strictly speaking this constraint is not actually necessary).
Conversely, an assignment making $a$ uniquely cheapest gives rise to a solution to the MILP with positive $\eps$.
For unary and binary cost functions, \cref{milp:local_necessity} is (essentially) equivalent to the formulation of Koster, van Hoesel, and Kolen~\cite[Section~6.3.2]{KosterVanHoeselKolen99Report}.

Of course, solving MILPs is \np-hard in general.
One might hope that this
specific MILP can be solved in polynomial time; however, this is not the case in
general: there is a simple reduction from \ksat[3]{} showing that, for a
suitable~$\Gamma$, deciding whether an available element is locally necessary at
a specific variable is \np-hard.
Thus, unless $\pclass=\np$, we cannot test local
necessity in polynomial time for every~$\Gamma$.
Hence, we work with the linear
programming (LP) relaxation of \cref{milp:local_necessity} instead.
Naturally, similar linear programming relaxations have been studied for CSPs and VCSPs before, e.g.,~\cite{Cooper10:osac,Kun12:itcs,KolmogorovThapperZivny15LP,crz22:sicomp}.

\begin{linearprogram}[Fractional local necessity test]
    \label{lp:fractional_local_necessity}
    Let $D$ be a finite domain and $\Gamma$ a finite-valued hard-list language over $D$.
    Let $I=(V,\CC,t)$ be an instance of \vcsp{}, and let $L_u$ be the nonempty set of available elements at each variable $u\in V$.
    Fix a variable $v\in V$ and an element $a\in L_v$.
    Let $N_v$ be the neighbors of $v$ in the primal graph, and let $\CC_v$ be the family of finite-valued constraints involving $v$.
    For each $C\in\CC_v$, let $Y_C$ be the product of the available sets of its other distinct variables, ordered as in \cref{milp:local_necessity}; with no other variables, it consists of the empty tuple.
    Relax \cref{milp:local_necessity} for $I$, $v$, and $a$ by replacing \cref{eq:local_necessity_neighbor_integrality,eq:local_necessity_tuple_integrality} with
    \[
        \begin{aligned}
            0\leq q_{u,d}\leq 1
             & \qquad\text{for every }u\in N_v,\ d\in L_u,       \\
            0\leq p_{C,\bar y}\leq 1
             & \qquad\text{for every }C\in\CC_v,\ \bar y\in Y_C.
        \end{aligned}
    \]
    All other constraints and the objective remain the same.
\end{linearprogram}

This gives us a fractional version of local necessity.

\begin{definition}[Locally fractionally necessary]
    \label{def:locally_fractionally_necessary}
    Let $D$ be a finite domain and $\Gamma$ a finite-valued hard-list language over $D$.
    Let $I$ be an instance of \vcsp{} in which every variable has a nonempty set of available elements.
    An element $a$ available at a variable $v$ of $I$ is \emph{locally fractionally necessary at $v$} if the optimum of \cref{lp:fractional_local_necessity} for $I$, $v$, and $a$ is positive.
\end{definition}

Clearly every feasible solution of the MILP is also feasible for its LP relaxation, with the same objective value.
Thus the LP optimum is at least the MILP optimum, and we obtain the following observation.

\begin{observation}
    \label{obs:fractional_necessity_safe}
    Every element that is not locally fractionally necessary is not locally necessary.
    Forbidding it at its variable using hard lists preserves the optimum and the primal graph.
\end{observation}

For unary and binary cost functions, our LP deletion criterion corresponds to the generalized Goldstein criterion~\cite{goldsteinEfficientRotamer1994}.
Lasters, De Maeyer, and Desmet~\cite{lastersEnhancedDeadEnd1995} apply this criterion in preprocessing using linear programming.
Koster, van Hoesel, and Kolen also use an LP relaxation in their approach~\cite[Section~8.3]{KosterVanHoeselKolen99Report}.
The higher-arity LP deletion criterion we use also exactly fits within Shekhovtsov's framework of relaxed improving mappings~\cite{shekhovtsovHigherOrderPersistency2016}.

The LP has rational coefficients, so exact rational linear programming determines its optimum in polynomial time.
This directly yields the following preprocessing step.

\begin{lemma}
    \label{thm:remove_fractionally_unnecessary}
    Let $D$ be a finite domain and $\Gamma$ a finite, finite-valued hard-list language over $D$.
    Given an instance $I$ of \vcsp{}, where the available set of each variable is nonempty, one can in polynomial time add hard-list constraints to obtain an instance $I'$ with the same optimum and primal graph such that every available element is locally fractionally necessary at its variable.
\end{lemma}
\begin{proof}
    While an available element is not locally fractionally necessary, forbid it at its variable by adding a hard list.
    By \cref{obs:fractional_necessity_safe}, each deletion preserves the optimum and the primal graph.
    The LP has polynomial size because the domain, language, and arities are fixed, and its rational coefficients have polynomial bit length.
    Exact rational linear programming determines whether its optimum is positive in polynomial time.
    The sole element of an available set of size one is locally fractionally necessary, so no available set becomes empty.
    At most $(|D|-1)|V|$ deletions occur, after which every available element is locally fractionally necessary.
\end{proof}

The LP deletion procedure subsumes the deletions in \cref{thm:remove_dominated_elements,thm:remove_noninteracting_pairs}, but we present these simpler steps first to expose the block structure and avoid running LPs on nonreversing blocks.

\subsubsection*{From Fractional Local Necessity to Expressive Sets.}

With the preprocessing above, we can already see how the final algorithm will work.
However, it remains to tie the size of the available sets that remain at the end of preprocessing to our notion of expressive sets.
In this section, the goal is to show precisely this: after all of our preprocessing, each remaining available set is expressive.

\begin{theorem}
    \label{thm:fractional_necessity_expressive}
    Let $D$ be a finite domain and $\Gamma$ a finite, finite-valued hard-list language over $D$, and $B$ a reversing block of $G_\Gamma$.
    Let $I$ be an instance of \vcsp{} such that the available set $L_v$ of each variable $v$:
    \begin{bracketenumerate}
        \item belongs to $B$ or has size one,
        \item is pairwise interacting, and
        \item has the property that each element of $L_v$ is locally fractionally necessary at $v$.
    \end{bracketenumerate}
    Then, every available set of $I$ is expressive.
\end{theorem}
\begin{proof}
    For each variable $u$ of $I$, write $L_u$ for its available set, and let $\mathcal A$ contain every nonempty subset of some $L_u$.
    By assumption, every member of $\mathcal A$ of size at least two is pairwise interacting and belongs to $B$.
    We first show that, for each $L\in\mathcal A$, there is an instance $J_L$ over $\Gamma$ with a variable $x$ whose available set is $L$, such that every variable of $J_L$ has an available set in $\mathcal A$ and every $a\in L$ is locally necessary at $x$ in $J_L$.

    \subparagraph*{Building instance $J_L$.}
    If $|L|=1$, take $J_L$ to have only the variable $x$ and a hard list on $L$.
    The claim holds because $x$ has only one available element.

    Now suppose $|L|\geq2$ and choose a variable $v$ of $I$ with $L\subseteq L_v$.
    In this case, for each $a \in L$, we use a solution to the LP that certifies that $a$ is locally fractionally necessary at $v$ to build the instance $J_L$.
    The basic idea is that although such a solution is fractional, we can create $J_L$ by using sufficiently many copies of the variables to clear the denominators.

    For each $a\in L$, local fractional necessity gives a rational solution of \cref{lp:fractional_local_necessity} for $I,v,a$ with margin (i.e., objective $\eps$ of the LP) $\eps_a>0$.
    Let $\CC_v$ be the family of finite-valued constraints involving $v$, and put $w_C=\weight{C}$ for $C\in\CC_v$.
    After identifying repeated variables and placing $v$ first, write $g_C(d,\bar y)$ for the cost of $C$ with $v$ assigned $d$, and let $Y_C$ be the product of the available sets of its other distinct variables, ordered as in \cref{lp:fractional_local_necessity}.
    Denote the assignments to the variables $p_{C, \bar y}$ in the chosen solution for $a$ by $p^a_{C,\bar y}$.
    For every $b\in L\setminus\{a\}$, \cref{eq:local_necessity_margin} of the LP gives
    \[
        \sum_{C\in\CC_v}w_C\sum_{\bar y\in Y_C}
        p^a_{C,\bar y}\bigl(g_C(b,\bar y)-g_C(a,\bar y)\bigr)
        \geq\eps_a>0.
    \]

    Choose a positive integer $N$ such that $Nw_C$ and $Nw_Cp^a_{C,\bar y}$ are integers for every $a$, $C$, and $\bar y$.
    Construct $J_L$ with a variable $x$ restricted to $L$ by a hard list.
    For each $C\in\CC_v$, add $Nw_C$ unit-weight copies of the original constraint $C$, using $x$ wherever $C$ uses $v$ and fresh variables for its other distinct variables in each copy.
    Preserve repeated occurrences of each variable within a copy, and give every fresh variable the available set of the corresponding variable of $I$ by a hard list.
    Thus every variable other than $x$ has an available set in $\mathcal A$.

    For each $a\in L$ and $C\in\CC_v$, assign the other variables in exactly $Nw_Cp^a_{C,\bar y}$ copies of $C$ the tuple $\bar y$, for each $\bar y\in Y_C$.
    These counts sum to $Nw_C$, and the variables in different copies are distinct, so this defines an available assignment $\pi^a$ to all variables of $J_L$ other than $x$.
    Changing $x$ from $a$ to any $b\in L\setminus\{a\}$ under this assignment changes the cost by
    \[
        N\sum_{C\in\CC_v}w_C\sum_{\bar y\in Y_C}
        p^a_{C,\bar y}\bigl(g_C(b,\bar y)-g_C(a,\bar y)\bigr)
        \geq N\eps_a>0.
    \]
    Hence every $a\in L$ is locally necessary at $x$ in this same instance $J_L$.

    \subparagraph*{Expressing cost functions.}
    We now prove by induction on $k\geq1$ that $\Gamma$ expresses all finite-valued unary cost functions on every $L\in\mathcal A$ with $|L|\leq k$.
    By \cref{cor:reversing_block_costs}, this also gives all nonnegative rational-valued cost functions, up to a constant, on products of such sets.

    If $|L| = 1$ any unary cost is constant, so the zero unary function suffices up to a constant.
    If $|L|=2$, interaction excludes domination in either direction, and \cref{thm:undominated_pair_unaries} gives all unary costs on $L$.
    This establishes the induction claim for $k\leq2$.

    Suppose the claim holds for sizes below $k\geq3$, and fix $L\in\mathcal A$ of size $k$ and $a_0\in L$.
    Take the instance $J_L$ above with variable $x$, and enumerate its other variables as $y_1,\dots,y_s$.
    For each $a\in L$, let $\bar y^a$ be their values under $\pi^a$.
    Write $F(d,\bar y)$ for the sum of the finite-valued constraint costs of $J_L$ when $x$ receives $d$ and the other variables receive $\bar y$.
    Then
    \[
        F(a,\bar y^a)<F(b,\bar y^a)
        \qquad\text{for every }a\in L\text{ and }b\in L\setminus\{a\}.
    \]
    The tuples $\bar y^a$ are pairwise distinct, since two distinct elements cannot both be uniquely cheapest under the same assignment; in particular, $s\geq1$.

    For $i\in[s]$, put
    \[
        T_i=\{\pi^a(y_i) \mid a\in L\setminus\{a_0\}\},
    \]
    and let $P=T_1\times\cdots\times T_s$.
    Each $T_i$ is a nonempty subset of the available set of $y_i$ in $J_L$, so $T_i\in\mathcal A$.
    Moreover, $|T_i|\leq k-1$ by its definition.
    Put
    \[
        K=\max_{d\in L,\ \bar y\in P}F(d,\bar y).
    \]
    Define the nonnegative function $\hat g :P\to\ratNnonneg$ by
    \[
        \hat g(\bar y)=
        \begin{cases}
            K-F(a,\bar y^a), & \bar y=\bar y^a
            \text{ for some }a\in L\setminus\{a_0\}, \\
            2K+1,            & \text{otherwise}.
        \end{cases}
    \]
    This function is well-defined since the tuples $\bar y^a$ and $\bar y^{a'}$ are distinct for $a \neq a'$.
    By the induction hypothesis and \cref{cor:reversing_block_costs}, there are a finite-valued $g\in\power{\Gamma}$ and a constant $c\in\ratN$ with $g(\bar y)=\hat g(\bar y)+c$ for all $\bar y\in P$.

    Keep the finite-valued constraints of $J_L$, restrict each $y_i$ to $T_i$ by a hard list, and add a gadget for $g$ with inputs $y_1,\dots,y_s$.
    Do not impose a hard list on $x$.
    Minimizing over all variables other than $x$ gives a unary function $h\in\power{\Gamma}$ that fulfills, for all $d \in L$,
    \[
        h(d)=\min_{\bar y\in P}\bigl(F(d,\bar y)+g(\bar y)\bigr).
    \]

    This function is finite-valued on all of $D$: $P$ is nonempty, and both $F$ and $g$ have finite values there.
    For $d\in L$, a tuple $\bar y\in P$ not among $\{\bar y^a \mid a \in L\setminus \{a_0\}\}$ has $F(d,\bar y)+\hat g(\bar y)\geq2K+1$, while every $\bar y^a$ with $a\neq a_0$ has $F(d,\bar y^a)+\hat g(\bar y^a)\leq2K$.
    Therefore
    \[
        h(d)=c+K+\min_{a\in L\setminus\{a_0\}}
        \bigl(F(d,\bar y^a)-F(a,\bar y^a)\bigr)
        \qquad(d\in L).
    \]

    We have that $\min_{a\in L\setminus\{a_0\}}
        \bigl(F(d,\bar y^a)-F(a,\bar y^a)\bigr)$ is zero when $d\in L\setminus\{a_0\}$, and is the positive rational number
    \[
        \delta=\min_{a\in L\setminus\{a_0\}}
        \bigl(F(a_0,\bar y^a)-F(a,\bar y^a)\bigr)
    \]
    when $d=a_0$.
    Scaling $h$ by $1/\delta$ thus gives, up to a constant shift, the unary cost that is one at $a_0$ and zero at every other element of $L$.
    Repeating this construction for each $a_0\in L$ and taking nonnegative weighted sums gives all finite-valued unary cost functions on $L$, up to a constant, completing the induction.

    For every available set $L_v$ of~$I$, the induction shows that $\Gamma$
    expresses all finite-valued unary cost functions on~$L_v$.
    Since $L_v$ is
    pairwise interacting by assumption, it is expressive by the remark after
    the restatement of \cref{def:expressive_set} at the start of \cref{sec:vcsp_hard_lists}.
\end{proof}

The preceding theorem lets us obtain an algorithm where the running-time base directly corresponds to the size of expressive sets belonging to reversing blocks.
Now, we are ready to provide our algorithm for finite-valued hard-list languages.

\begin{theorem}
    \label{thm:hard_lists_algorithm}
    Let $D$ be a fixed finite domain and $\Gamma$ a fixed finite, finite-valued hard-list language over $D$.
    Given an instance $I$ of \vcsp{} and a tree decomposition $\mathcal T$ of its primal graph of width $\tw$, one can compute $\Opt(I)$ and an assignment attaining it in time
    \[
        (c_{\Gamma})^{\tw}\cdot (|I|+|\mathcal T|)^{\Oh(1)}.
    \]
\end{theorem}
\begin{proof}
    If some variable has no available element, then $I$ has no finite-cost
    assignment, so return the value $\infty$ with an arbitrary variable assignment.
    Otherwise apply \cref{thm:remove_noninteracting_pairs} to obtain nonempty, pairwise interacting available sets.
    The optimum and primal graph are preserved.
    By \cref{thm:independent_block_instances}, separate the instance into block instances, recording their additive constant and transforming the supplied tree decomposition with at most a constant increase in width.
    The block instances use the original language $\Gamma$, and their auxiliary variables have available sets of size one.
    Compute the optimum and an optimal assignment for every nonreversing block
    in polynomial time by \cref{thm:solve_nonreversing_block}.

    For each reversing block $B$, apply \cref{thm:remove_fractionally_unnecessary} to its block instance.
    All remaining available sets are still pairwise interacting, and each set of size at least two still belongs to $B$, since these properties are inherited by subsets.
    By \cref{thm:fractional_necessity_expressive}, every remaining available set is expressive and therefore has size at most $c_\Gamma$ by \cref{thm:hard_lists_base_equality}.
    The LP step only adds hard-list constraints, so the primal treewidth remains bounded by $\tw+\Oh(1)$.

    Using \cref{thm:available_sets_dp}, compute the optimum and an optimal assignment of each reversing block instance in time $(c_\Gamma)^{\tw}\cdot (|I|+|\mathcal T|)^{\Oh(1)}$, where we use that $c_\Gamma$ is a fixed constant.
    There are only constantly many blocks, since $D$ and $\Gamma$ are fixed.
    Add their optimum values and the recorded constant, and combine their optimal assignments with the original singleton values as in \cref{thm:independent_block_instances}.
\end{proof}

\subsection{The Main Theorem}

Now, we are ready for the proof of \cref{thm:main}.

\mainTheorem*
\begin{proof}
    We obtain statement (1) from \cref{thm:hard_lists_algorithm}.
    For (2), assume that \twseth{} (\pwseth{}, respectively) holds and that $c_\Gamma \geq 2$.
    Then, \cref{thm:lb_from_c_gamma} gives (2).
\end{proof}

\section{AI Usage}
\label{sec:ai_usage}

The proof of the main result of this paper has been obtained from several conversations with GPT-5.6 Sol and GPT-6 Astra based on our notes and partial results.
The draft has been written with the help of various large language models (also including GPT-6.1 Sol), which
were also used for the purpose of literature search.
The introduction and technical overview were written by the authors.
The content has been checked and curated by humans, and the authors take full responsibility for the content of the article.

\bibliographystyle{alphaurl}
\bibliography{bib}

\clearpage

\appendix
\crefalias{section}{appendix}

\section{Restricting the Domain of \texorpdfstring{$\Gamma$ vs.~$\power{\Gamma}$}{Γ vs. pow(Γ)}}
\label{sec:gamma_pow_gamma_example}

\cref{thm:main} shows that the complexity of the problem is determined by the maximum size $c_\Gamma$ of a set $S\subseteq D$ on which $\power{\Gamma}$ contains a disequality function. In a sense, this means that $\Gamma$ has a ``core'' $S$ on which VCSP is as hard as possible, i.e., requires $|S|^\tw\cdot n^{O(1)}$ time. The following example clarifies that this \emph{does not} necessarily mean that if we restrict every relation in $\Gamma$ to $S$, then we get a language $\Gamma_S$ with the same complexity, also requiring $|S|^\tw\cdot n^{O(1)}$ time. It is possible that in order to express the disequality function on $S$ appearing in  $\power{\Gamma}$, we need to rely on values outside $S$.
\begin{example}
    Let $D=\{0,1,2\}$, and let $\Gamma$ be the finite-valued hard-list language consisting of all hard lists and the binary cost function $f(x,y)$ defined by the table
    \[
        \begin{array}{c|ccc}
            f & 0 & 1 & 2 \\\hline
            0 & 0 & 0 & 0 \\
            1 & 1 & 2 & 0 \\
            2 & 2 & 1 & 0
        \end{array},
    \]
    where rows give the first argument and columns the second.
    Using hard lists we can obtain the unary cost functions $f(\cdot,0)=(0,1,2)$, $f(\cdot,1)=(0,2,1)$, and $f(2,\cdot)=(2,1,0)$.
    Here each triple gives the function's costs at the domain values $0,1,2$, respectively.
    These provide strict preferences in both directions on every pair, so no distinct values dominate one another.
    Each pair of domain elements also witnesses its own interaction through $f$, so all three pairs belong to one block by \cref{thm:available_set_block}.

    Nevertheless, for every $c\in D$ we have
    \[
        f(1,c)\geq\frac{f(0,c)+f(2,c)}2,
        \qquad
        f(c,1)\geq\frac{f(c,0)+f(c,2)}2.
    \]
    Moreover, $f(1,1)=2\geq 0=(f(0,0)+f(2,2))/2$ as well.

    Now, consider an arbitrary instance of the problem.
    Let $v$ be a variable whose available set is $D$, and let $h_v(d,\pi)$ be the total cost of the constraints that contain $v$ when assigning $d$ to $v$ and $\pi$ to the neighbors of $v$ in the primal graph.
    Summing the inequalities gives
    \[
        h_v(1,\pi)\geq\frac{h_v(0,\pi)+h_v(2,\pi)}2.
    \]
    Thus either $0$ or $2$ can replace $1$ without increasing the cost, although the choice may depend on $\pi$.
    In fact, when any variable of any instance has all of $D$ as its available set, then the element $1$ is not locally fractionally necessary.
    The preprocessing therefore leaves available sets of size at most two, so \cref{thm:available_sets_dp} gives running time $2^{\tw}\cdot (|I|+|\mathcal T|)^{\Oh(1)}$ for an instance $I$ with a supplied tree decomposition $\mathcal T$ of width $\tw$.

    For a nonempty proper subset $S\subsetneq D$, let $\Gamma_S$ be obtained by restricting every function of $\Gamma$ to $S$.
    Every problem \vcsp[\Gamma_S]{} is polynomial-time solvable.
    On $\{0,1\}$, value $0$ dominates $1$; on $\{1,2\}$, value $2$ dominates $1$.
    Deleting the dominated value whenever both are available fixes every variable.
    On $\{0,2\}$, the only positive entry is $f(2,0)=2$, so the problem is a weighted directed minimum-cut problem.
    The hard lists force some vertices to be on a specific side of the cut, which is easy to handle.

    In contrast, $\power{\Gamma}$ contains the disequality function on $\{0,1\}$.
    To see this, let $\ell_S$ denote the hard-list cost function that is $0$ on $S$ and $\infty$ outside $S$, and define
    \[
        g(x,y)=\ell_{\{0,1\}}(x)+\ell_{\{0,1\}}(y)
        +\min_{p\in D}\bigl(2f(x,y)+3f(p,x)+f(p,y)+\ell_{\{2\}}(p)\bigr)-7.
    \]
    The hard list forces $p=2$, and the weighted sum is $8$ at $(x,y)=(0,0),(1,1)$ and $7$ at $(x,y)=(0,1),(1,0)$.
    Thus $g(x,x)=1$ for $x\in\{0,1\}$ and $g(x,y)=0$ for distinct $x,y\in\{0,1\}$, while $g$ is infinite outside $\{0,1\}^2$.
    In particular, $g$ is nonnegative, so the additive shift is allowed and $g\in\power{\Gamma}$.
    Hence $c_\Gamma\geq 2$, and \cref{thm:main} rules out an algorithm for \vcsp[\Gamma]{} running in time $(2-\eps)^\tw\cdot (|I|+|\mathcal T|)^{\Oh(1)}$ for any $\eps>0$, unless \twseth{} is false.
    This matches the upper bound above, even though every proper domain restriction is polynomial-time solvable: the gadget expressing disequality on $\{0,1\}$ uses the auxiliary value $2$.
    The same gadget also gives \np-hardness by a reduction from \problemname{Max Cut}~\cite{gareySimplifiedNPComplete1976}, since placing it on every edge of a graph $G$ gives total cost $8|E(G)|$ minus the cut size before the additive shifts.
    Therefore, unless $\pclass=\np$, the full problem cannot reduce in polynomial time to any single proper domain restriction $\Gamma_S$.
\end{example}

\end{document}